\documentclass{aa} 
\usepackage{graphicx}
\begin{document}
\newcommand{\kms}{{\,     \rm     km~s}^{-1}}    \newcommand{\ho}{{\rm
km~s}^{-1}~{\rm Mpc}^{-1}} \newcommand{\obj}{HE~1434$-$1600}
\title{On-axis spectroscopy  of the $z=0.144$  radio-loud quasar \obj:
an  elliptical host with a highly ionized ISM
\thanks{Based on  observations made with the ESO  Very Large Telescope
ANTU/UT1 at ESO-Paranal observatory, Chile (program 65.P-0361(A)).}}

\author{G.  Letawe  \inst{1} \and F.  Courbin \inst{1}  \and P. Magain
\inst{1}  \and M.   Hilker \inst{2}  \and P.   Jablonka  \inst{3} \and
K. Jahnke \inst{4} \and L. Wisotzki \inst{4,5} }

\offprints{G.  Letawe}  

\institute{Institut d'Astrophysique et de G\'eophysique, Universit\'e de  
Li\`ege, All\'ee du 6  Ao\^ut, 17, Bat.  B5C, Li\`ege 1, Belgium
\and  Sternwarte der  Universit\"at Bonn,  Auf dem  H\"ugel  71, 53121
Bonn, Germany
\and GEPI, Observatoire de  Paris, Place Jules Janssen, F-92915 Meudon
Cedex, France
\and Astrophysikalisches Institut Potsdam, An der Sternwarte 16, 
14482, Potsdam, Germany
\and Institut f\"ur Physik, Universit\"at Potsdam, Am Neuen Palais 10,
14469 Potsdam, Germany }

\date{}

\abstract{VLT on-axis optical spectroscopy of the $z=0.144$ radio-loud
quasar \obj\ is presented. The  spatially resolved spectra of the host
galaxy are deconvolved and separated  from those of the central quasar
in order  to study the dynamics  of the stars  and gas as well  as the
physical conditions of the ISM.  We  find that the host of \obj\ is an
elliptical  galaxy  that resides  in  a group  of  at  least 5  member
galaxies, and that most likely experienced a recent collision with its
nearest companion. Compared with other quasar host galaxies, \obj\ has
a highly  ionized ISM.   The ionization state  corresponds to  that of
typical Seyferts,  but the ionized  regions are not distributed  in a
homogeneous way around the QSO, and are located preferentially several
kiloparsecs away  from it. While  the stellar absorption lines  do not
show any  significant velocity field,  the gas emission lines  do. The
observed gas velocity field is hard to reconcile with dynamical models
involving rotating disk, modified Hubble  laws or power laws, that all
require  extreme central  masses ($M>10^{9}$~M$_{\odot}$)  to provide
only poor fit to the data.  Power law models, which best fit the data,
provide a  total mass of $M(<$10~kpc$)=9.2\,10^{10}$~M$_{\odot}$.  We
conclude  that the recent  interaction between  \obj\ and  its closest
companion has strongly affected the gas velocity and ionization state,
from  the   center  of  the   galaxy  to  its  most   external  parts.
\keywords{Galaxies:  dynamics, interaction,  black  holes --  quasars:
individual:  HE~1434-1600, quasar host  galaxies --  techniques: image
processing, deconvolution}}

\titlerunning{On-axis   spectroscopy   of   quasar   hosts   galaxies:
HE~1434-1600}

\maketitle


\section{Introduction}

Spectroscopy  of quasar  host galaxies,  either  with a  long slit  or
through the use of integral field spectrographs now available on large
telescopes, is able to provide unique clues on the stellar content, on
the physics  of the interstellar medium  (ISM) and on  the dynamics of
these complex objects.  Since their otherwise luminous active galactic
nucleus (AGN) is extinguished by the central molecular torus, the  radio galaxies can be rather easily  studied, even in
spectroscopy (e.g., Vernet et al.  \cite{vernet01}). This is also true
for Seyfert galaxies  where the brightness of the  central AGN is much
lower  than  in genuine  quasars.   On  the  contrary, obtaining  high
quality  spectra of the  host of  bright quasars  remains a  much more
challenging task  because of the high luminosity  contrast between the
central AGN and the host.

We have  carried out a  systematic spectroscopic campaign of  the host
galaxies of a sample of  bright quasars, using the European Very Large
Telescope (VLT), at Paranal  observatory, Chile.  The quasar sample is
selected  from   the  Hamburg-ESO   Survey  (HES;  Wisotzki   et  al.\
\cite{wisotzki00}).  It includes  20 intrinsically bright quasars with
$M_{B}  < -23$  and  $z  < 0.33$.   No  prior morphological  selection
criteria have been  applied to the sample. The  full sample, for which
optical and  near-IR imaging is also  available (Jahnke \cite{jahnke};
Jahnke et al.\ \cite{jahnke03, jahnke04}) will be fully described and
analyzed in Letawe et  al., in prep., while the deconvolution
techniques used to decompose the  data into two independent spectra of
the  unresolved central  quasar and  of the  extended host  galaxy are
presented  in  Magain  et  al.   (\cite{Magain}) and  Courbin  et  al.
(\cite{Courbin_a}).  A  first application  of these techniques  to the
quasar    HE~1503+0228    is    presented    in   Courbin    et    al.
(\cite{Courbin_b}).  It shows that the host of this quasar is a normal
spiral  galaxy  with $M(r<10$~kpc$)  =  (1.9  \pm  0.3) \times  10^{11}~$M$_{\odot}$  and with stellar  populations typical  of a  normal spiral
galaxy.

Some of  the objects  in our sample  display peculiar  features. \obj,
with $z=0.144$ and absolute $B$ magnitude of $-24.3$, stands indeed as
a   rather   special  case,   which   appeared   to  deserve   further
investigations.  The peculiar velocity field, prominent emission lines
and rich environment,  led us  to obtain more data for this radio-loud
quasar than for  the rest of the sample.   The present paper describes
the observations  of what appears to  be an elliptical  galaxy that is
probably  experiencing  a collision,  and  that  shows a  particularly
strongly ionized Inter Stellar Medium (ISM).

All along  this paper,  the conversion of  angular scales  into linear
distances is done using $H_0=65~\ho $, $\Omega_m=0.3$  and  $\Omega_{\Lambda}=0.7$ leading to
a scale of 2.73~kpc per arcsecond at $z=0.144$.


\section{Imaging}

\label{img}
\begin{figure}
\centering \includegraphics[width=8.6cm,height=6.5cm]{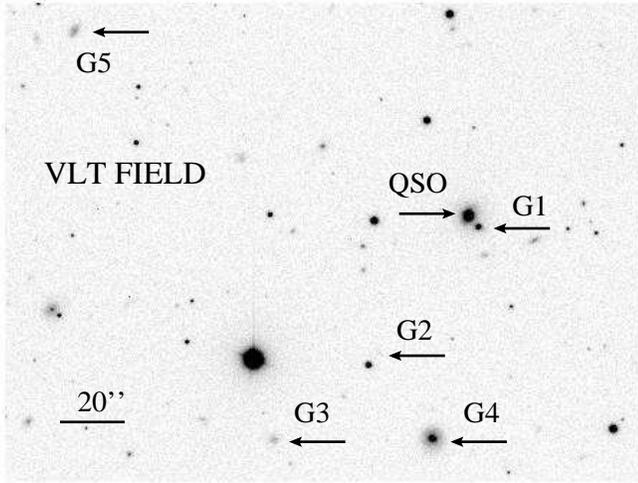}
\caption{Part of the  FORS1 field of view around  \obj.  Five galaxies
with measured redshifts are  indicated by horizontal arrows.  They all
have redshifts very  similar to that of the  quasar, $z\sim 0.144$ (see
Section  \ref{neighb}). North is  up and  East is  left, as  all other
figures in this paper.}
\label{field}
\end{figure}

\subsection{Optical $-$ near-IR}

Optical, near-IR and radio images of \obj\ were available, in addition
to the spectroscopic  data.  Two VLT $R$-band images  were taken prior
to each  spectroscopic observation in  order to construct  the slitlet
mask.  These images were obtained on  the nights of April 11, 2000 (30
sec exposure,  seeing 0\farcs7) with FORS1  and on the  night of May
13, 2002 (15~sec, seeing 1\farcs1) with FORS2.   Fig.~1 shows part
of the  FORS1 field  of view. Additional  images were obtained  in the
$V$-band at the 1.54m Danish  telescope at La Silla observatory and in
the infrared $J$-band with the ESO 3.5m New Technology Telescope (NTT)
equipped with the instrument SOFI.  The $V$- and $J$-band data are the
same as used by Jahnke \cite{jahnke04}.  The technical details of the
observations are given in Table \ref{obs}.

\begin{figure}[t]
\centering                     
\includegraphics[width=4.3cm]{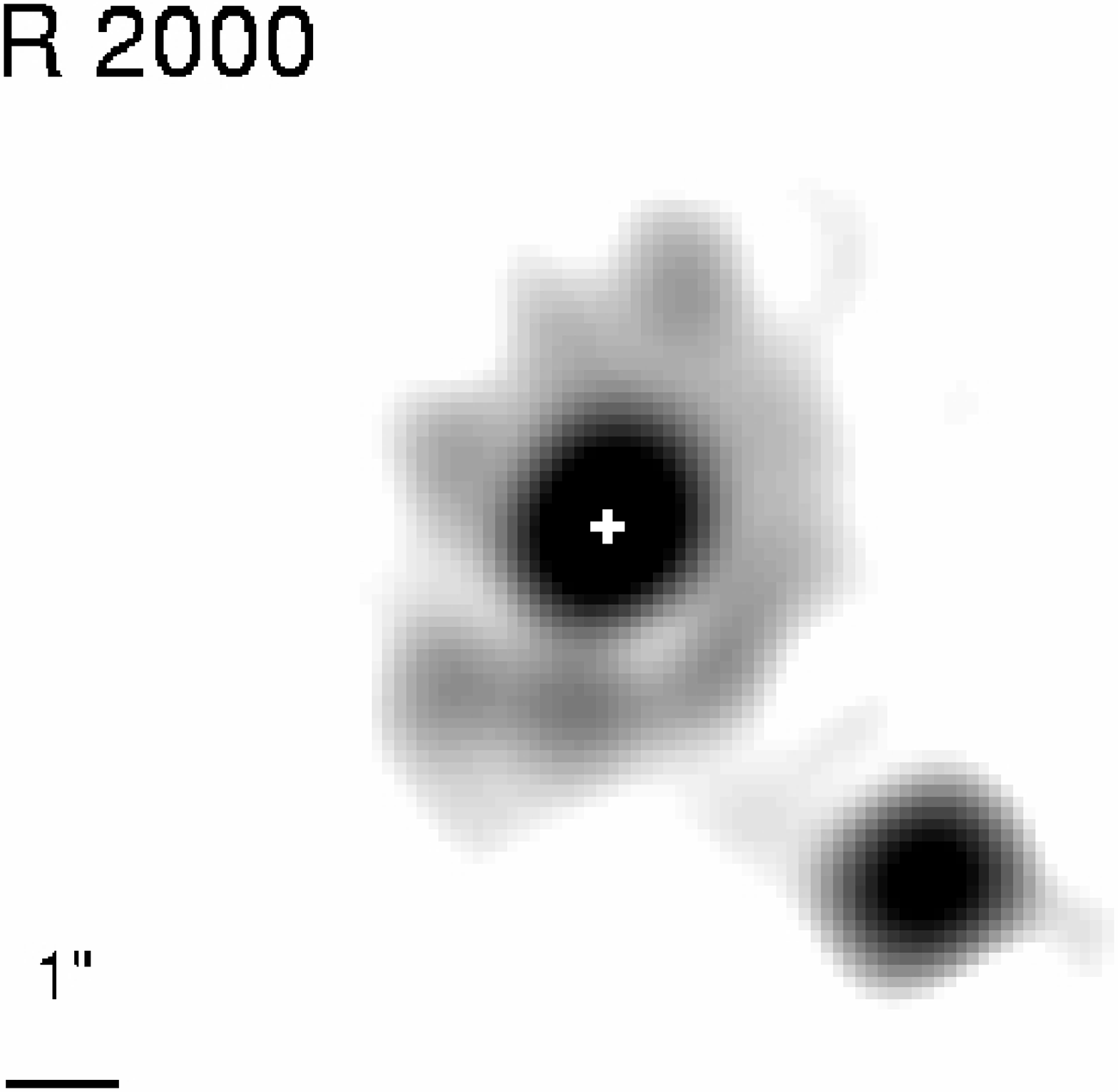}
\includegraphics[width=4.3cm]{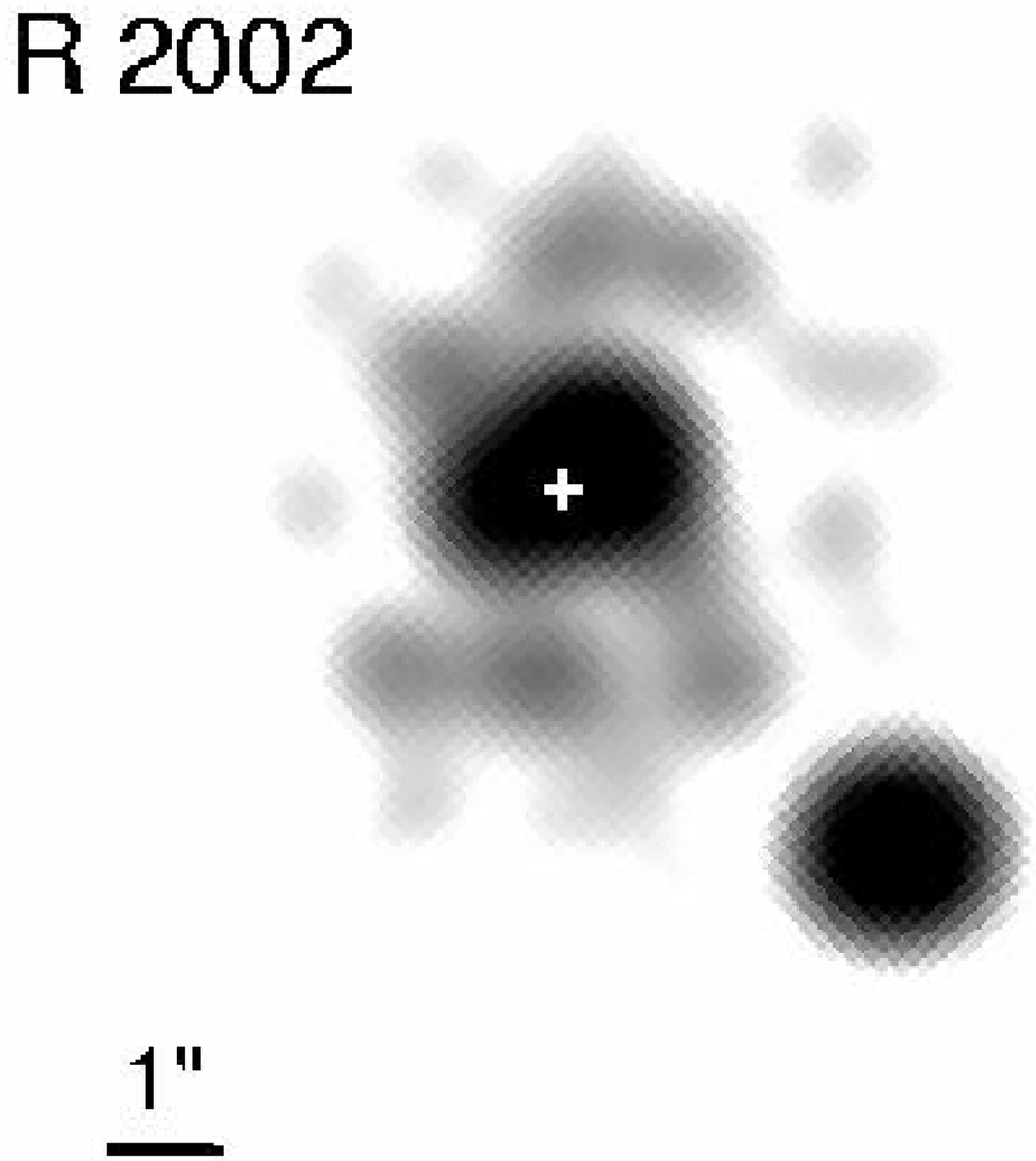}
\includegraphics[width=4.3cm]{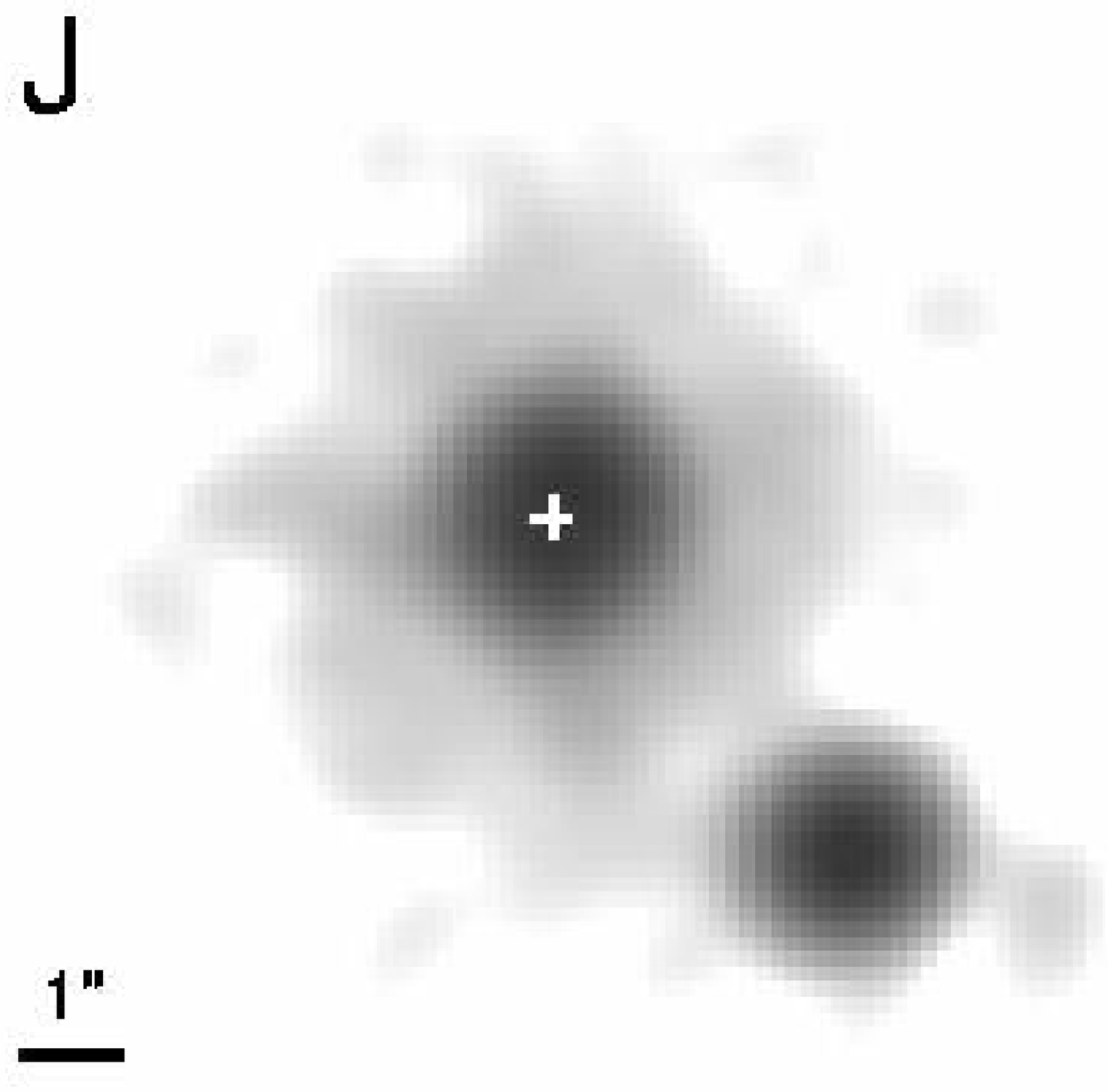}
\caption{Deconvolved images of the  host galaxy alone, with the quasar
removed. The  position of the quasar  is indicated by  the cross.  The
intensity scale  is different  in each image,  as well as  the angular
resolutions. While the $J$-band  image appears smooth, both $R$ images
show  structure about  2\arcsec\  away  from the  center  of the  host
galaxy.}
\label{images} 
\end{figure}

\begin{table}[b]
\centering
\caption{Details of the imaging  observations.  The first column gives
the telescope  and instrument used, column  4 gives the  pixel size in
arcseconds,  while  the  last  column  gives the  integration  time  in
seconds.}
\begin{tabular}{ccccc}
\hline\hline Instr.  &  Band & Date & scale&  $t$\\ 
\hline VLT-FORS2 &$R$ & 13-05-2002 &0.252 & 15s \\  
VLT-FORS1 & $R$ & 11-04-2000 & 0.200 &30s \\  
NTT-SOFI & $J$ &  24-02-1999&0.290 & 300s \\  
1.54m-DFOSC& $V$ &14-04-1997& 0.390 & 600s\\ 
\hline
\end{tabular}
\label{obs}
\end{table}

\begin{figure}[t]
\centering \includegraphics[width=9cm,angle=-90]{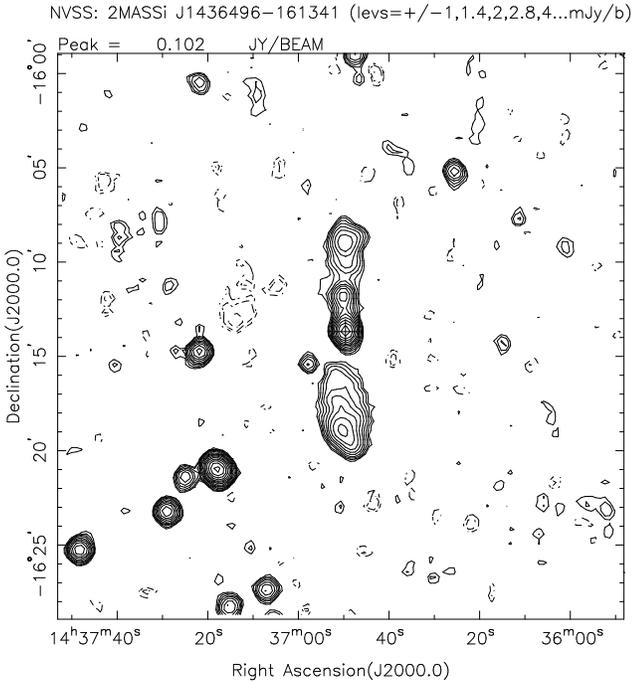}
\caption{NVSS image  centered on  the position of  \obj. The  field of
view is 30\arcmin$\times$30\arcmin, and the beam size in this 1.4~GHz  map  is  45\arcsec.   The  central ``hot  spot''  corresponds  to
\obj. Two lobes also extend over  5--6\arcmin to the north and south of
the  quasar.}
\label{nvss}
\end{figure}

All images, except  the low S/N $V$-band image  were deconvolved using
the MCS  algorithm (Magain, Courbin \& Sohy  \cite{Magain}), in order
to separate  the quasar  from its host  galaxy and to  reveal possible
substructures that may help to explain the spectroscopic observations.
The  deconvolved host  images  all show  extended  and rather  uniform
emission (Fig.~\ref{images}),  with a significant bulge contribution.
There   are,  however,   some  deviations   to  the   smooth  profile,
2--5\arcsec\, away  from the nucleus.  These  small scale substructures
match well  the ones  seen in [\ion{O}{iii}]  in the  spectroscopic observations
(see Section \ref{obsspec}).

A close companion,  labeled G1 on Fig.~\ref{field}, can  be seen next
to  \obj,  located 4\arcsec\,  from  the  center,  or 11~kpc.   Its
redshift (see  Section~\ref{neighb}) is the  same as that  of \obj.  A
$r^{-1/4}$  law   fits  well  its  1D  intensity   profile,  while  an
exponential disk  does not.   Moreover, its shape  is that  of an
elliptical galaxy.

Photometry of  the quasar  host has been  performed by Jahnke  et al.\
(\cite{jahnke04}) in a multicolour  study of QSO hosts. After masking
of the close companion they were  able to model the host of \obj\ with
a pure  spheroid light distribution.   The measured magnitudes  of the
host are $V=17.0$, $R=16.4$, $J=15.1$, and $K=13.7$.  They corrected for galactic extinction using the maps by Schlegel et al. (\cite{schlegel98}) and applied a K-correction term that was computed from a measured
broad-band SED (spectral energy distribution). For that purpose, template SEDs were fitted to six photometric points in the optical and NIR and  the template was used to derive a K-correction term
(see Jahnke et al., \cite{jahnke04} for details). Including these
corrections the colours of the host are $V-R=0.43$, $V-J=1.66$ and
$V-K=2.75$, i.e.   significantly  bluer colours  than
expected  for  an inactive  galaxy  of  this  luminosity (Fukugita  et
al. 1995,  Fioc \& Rocca-Volmerange  1999).  Modeling of  the stellar
populations  points  to  either  an  overall  young  dominant  stellar
population of  1--2~Gyr, or may  also be indicative of  continuous star
formation, plus  a small fraction (1\,\%  in mass) of  a recent (e.g.,
100~Myr) star burst population.

\subsection{Radio map}

A radio  map of \obj\  has also been  retrieved from the NRAO  VLA Sky
Survey  (Condon   et  al.   \cite{nvss}),  through   the  online  NASA
Extragalactic  Database (NED). It  shows large  lobes extending  up to
6\arcmin\ or  980 kpc, North and South of the quasar (Fig.~\ref{nvss}).
With an integrated  radio flux of 7.14~mJy at 1.4~GHz and a magnitude
$B=15.62$   (quasar  plus   host),  we   estimate  a   radio  loudness
$R=F_{\mathrm{5\,GHz}}$/$F_{4400}\sim 10$, a  value which places \obj\ at
the  limit  between radio-loud  and  radio-quiet quasars  (Kellermann 
\cite{kellermann}).  Such  a radio structure has  generally been found
to be associated  with extended optical emission lines  for example in
radio galaxies (e.g., Fosbury et al. \cite{fosb}), Seyferts (Morganti
et al.  \cite{morganti}) or in  Radio Loud Quasars (RLQ)  (Boroson et
al.   \cite{boroson84},  \cite{boroson85};   Stockton   \&  MacKenty 
\cite{stock87}). We investigate  in Section~\ref{ISM}, the possibility
that \obj\ is  a radio loud quasar (although  almost radio quiet) with
jet induced ionization of the ISM.

\section{Spectroscopy: observations and data reduction}
\label{obsspec}

Our first spectroscopic observations were obtained during the night of
April 11,  2000, with the FOcal Reducer/low  dispersion Spectrograph 1
(FORS1) mounted on  the 8.2m VLT/UT1 ANTU, at  ESO-Paranal.  Since the
aim  of  the project  is  to  spatially  deconvolve the  spectra,  the
Multi-Object-Spectroscopy (MOS) mode was  chosen in order to allow the
simultaneous observation of the  quasar and of neighbouring stars used
to determine  the Point Spread  Function (PSF). The same  approach was
used in Courbin et al.\  (\cite{Courbin_b}) in order to study the host
of HE~1503+0228, at $z=0.135$. Since  19 slitlets are available, we were
also able to observe several galaxies in the vicinity of the quasar.

\begin{figure}[t!]
\centering \includegraphics[width=8cm]{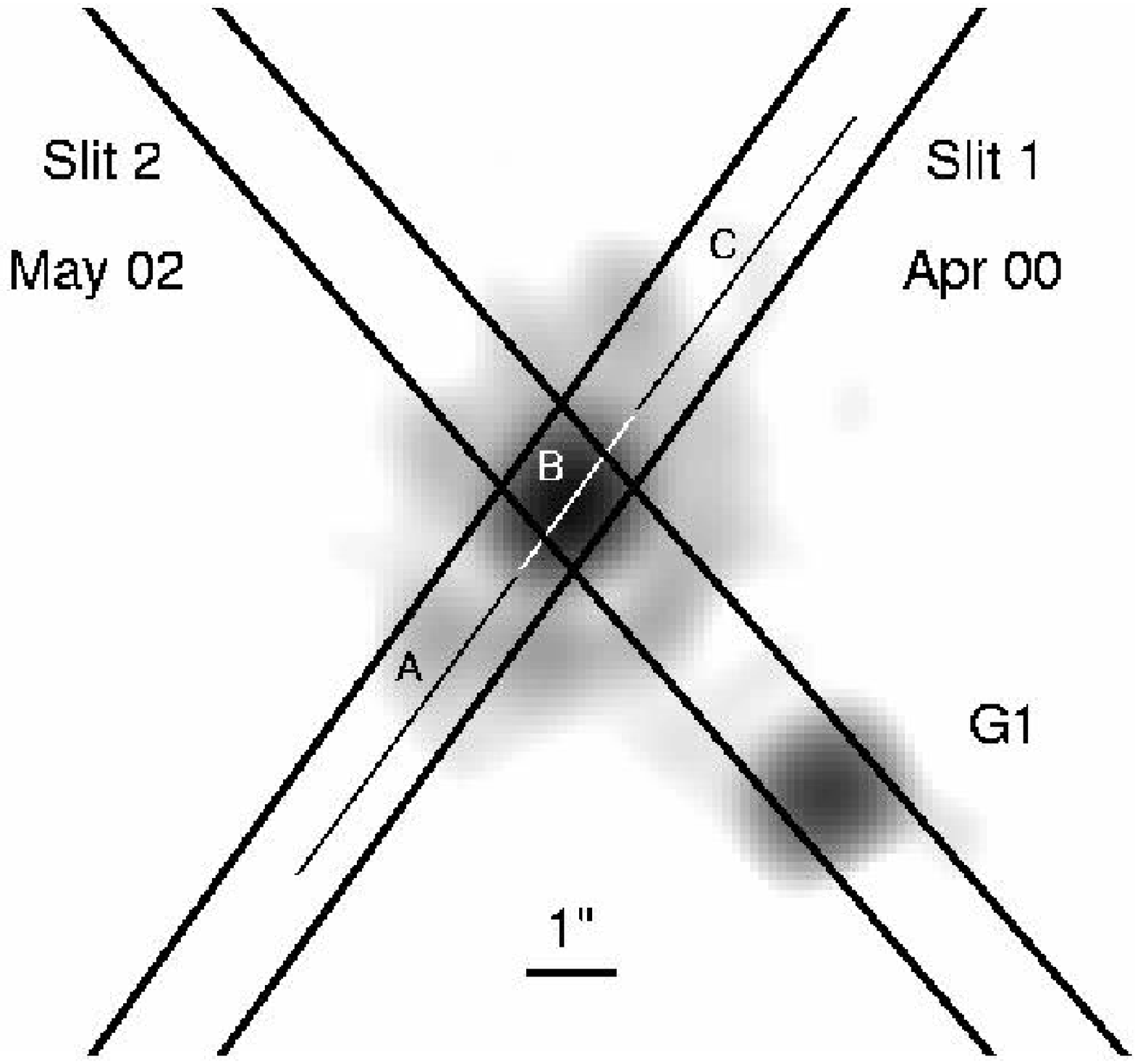}
\caption{Slit orientations  represented in overlay  on the deconvolved
$R$-band image of the host, after removal of the quasar. The companion
galaxy G1  is also visible on  this image.  Slit~1  corresponds to the
data  taken in  April 2000,  and  slit~2  corresponds to  the May  2002
data. In  each case the  slit width  is 1\arcsec. A,  B and C  are the
three   regions  we  used   to  extract   the  spectra   (see  Section
\ref{obsspec}). The total size of this  image is about the size of the
central hot-spot in the radio map of Fig.~\ref{nvss}.}
\label{slits}
\end{figure}

\begin{figure}[h!]
\centering 
\includegraphics[width=4cm]{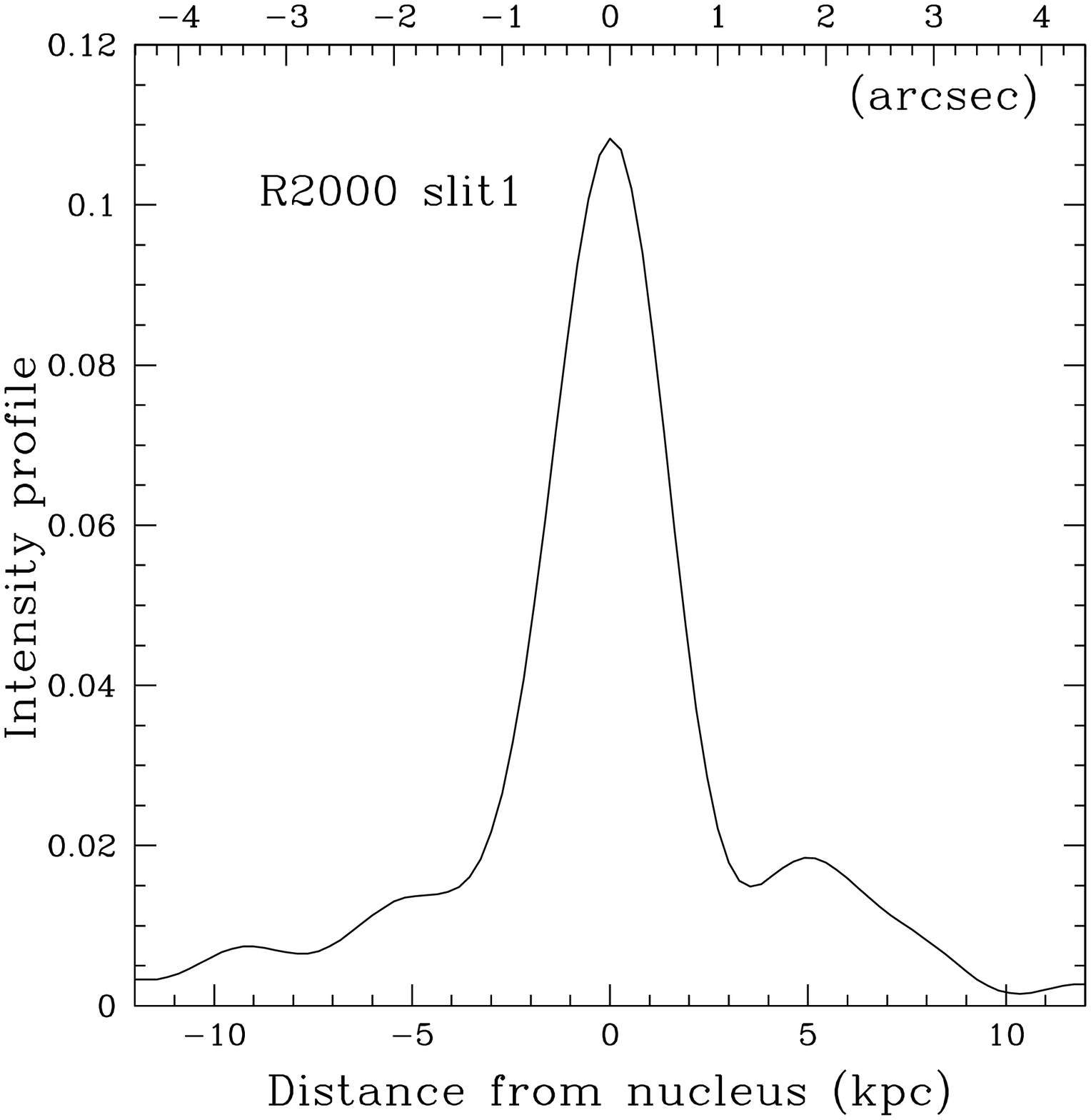}
\includegraphics[width=4cm]{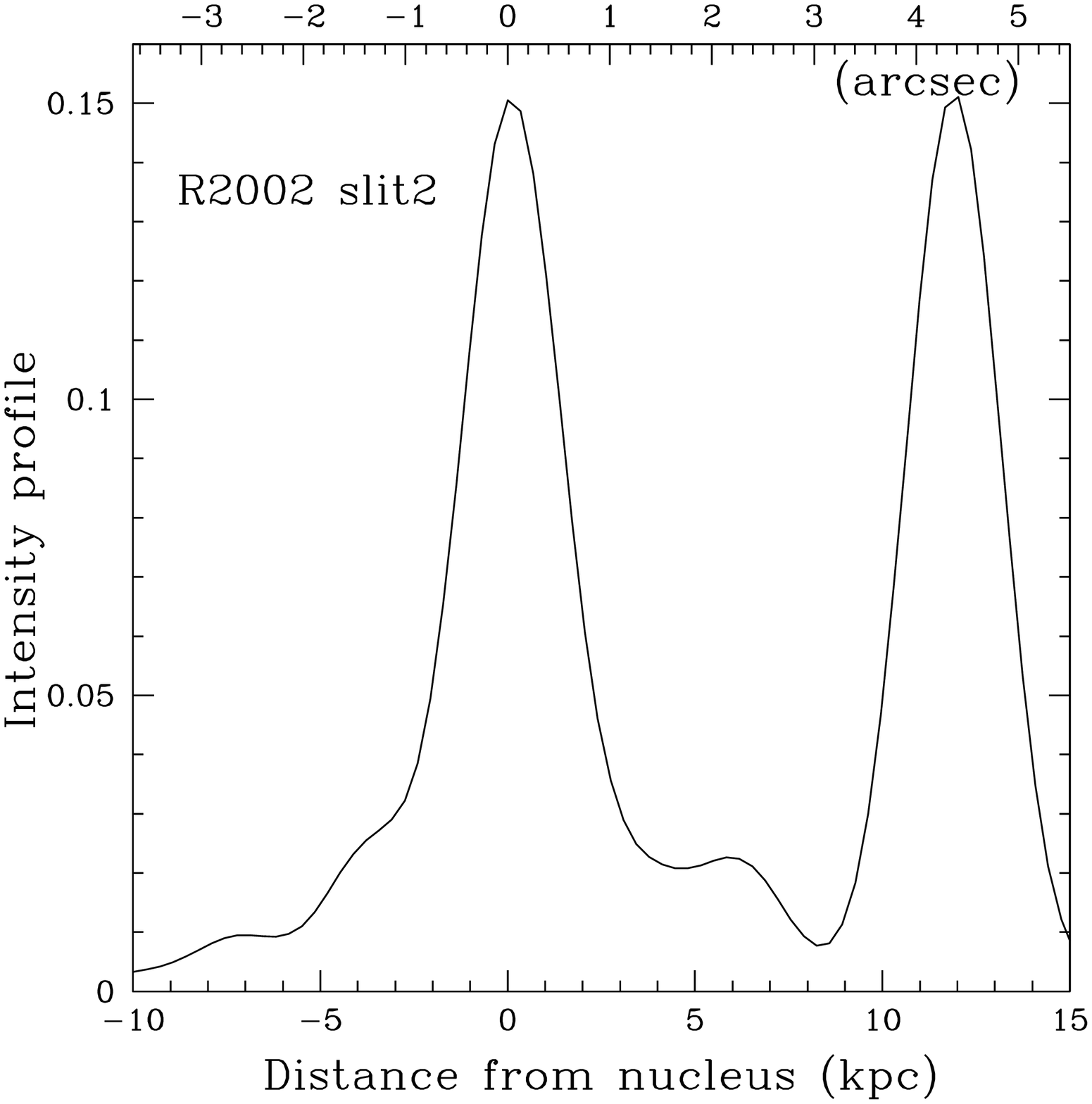}
\caption{Extracted  intensity  profiles  along  {\it slit1}  and  {\it
slit~2}  of the  host  galaxy, as  defined  in Fig.~\ref{slits}.   Each
profile is obtained by  spatially integrating the deconvolved $R$-band
image  across  the 2  slits,  after  removal  of the  quasar.  Obvious
structures with a typical size of  1~kpc do appear 2\arcsec\ away from
the quasar.}
\label{extracted_profiles}
\end{figure}

Three grisms were  used (G600B, G600R, G600I), with  a resolving power
$R \sim 700$.   The exposure time for each grism  was 1200 seconds and
the full spectrum  covers the wavelength range between  3250~\AA\ and
8000~\AA\ in the rest frame.   All basic reduction steps were made with
the IRAF package, leading to flatfielded, flux calibrated and rebinned
spectra with 0\farcs2\ per pixel  in the spatial direction and 1~\AA\
per pixel  in the  spectral direction.  Since  there is  a significant
slit curvature  in some of  the slitlets, a 2D  wavelength calibration
was performed,  as well  as a 2D  sky subtraction.  The  seeing varied
between 0\farcs5  \, and 0\farcs6.  The slit  orientation (PA) was
35$^{\circ}$ clockwise from north. We refer to these data as {\it slit
1} all along the text and in Fig.~\ref{slits}.

\begin{figure*}[p]
\centering \includegraphics[width=15cm,height=10cm]{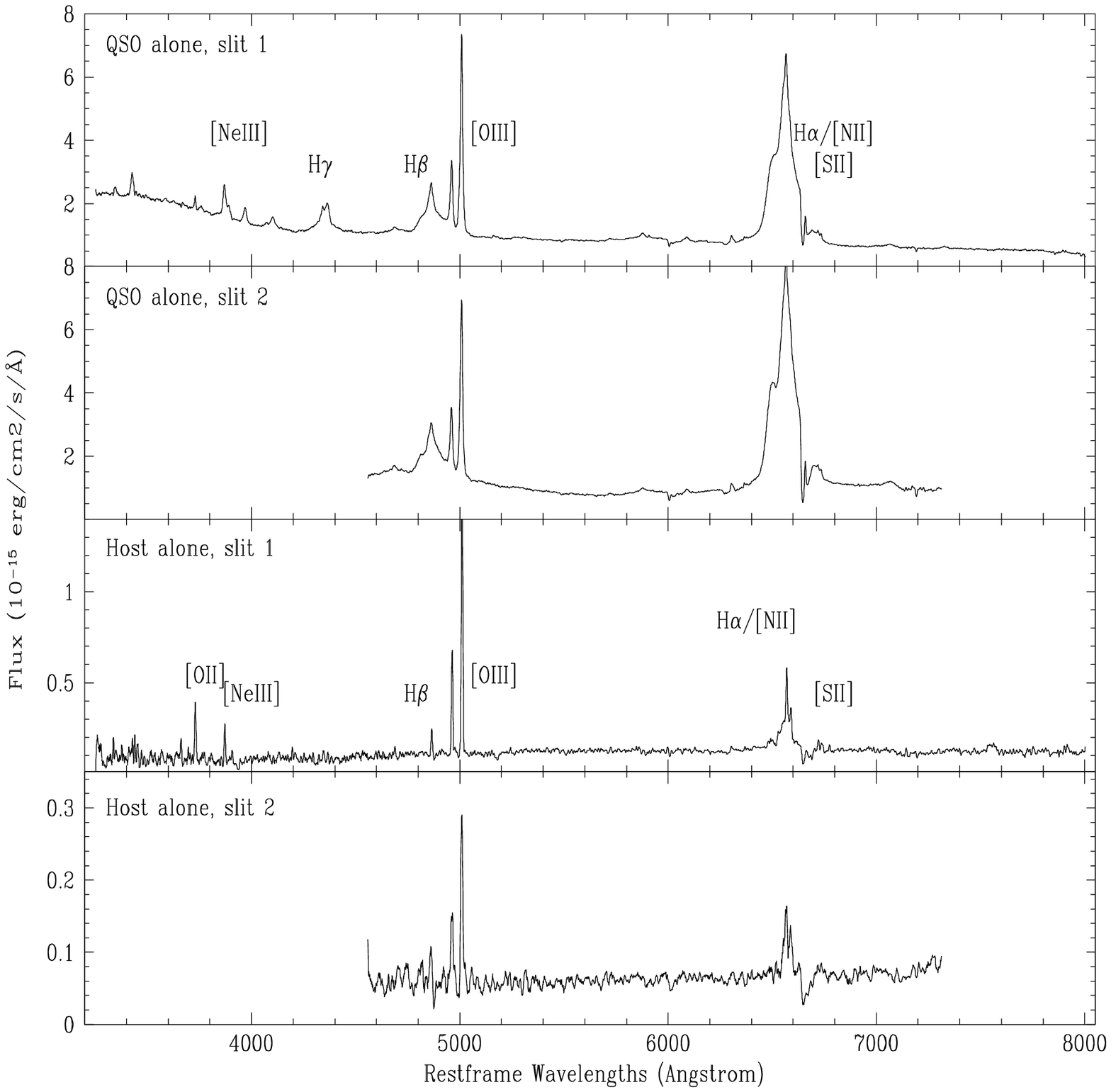}
\caption{1D  spectra   of  the  quasar  and  the   host  galaxy  after
deconvolution.   The   spectrum  of   the  host  is   integrated  over
6\arcsec. All  spectra are displayed  in the rest frame, for  the two
slit  orientations. The  quasar spectra  are  shown in  the two  upper
panels, while the  spectra of the host galaxy alone,  are shown in the
two lower ones. Note the different flux scales.}
\label{spec1D}
\centering \includegraphics[width=15cm,height=11cm]{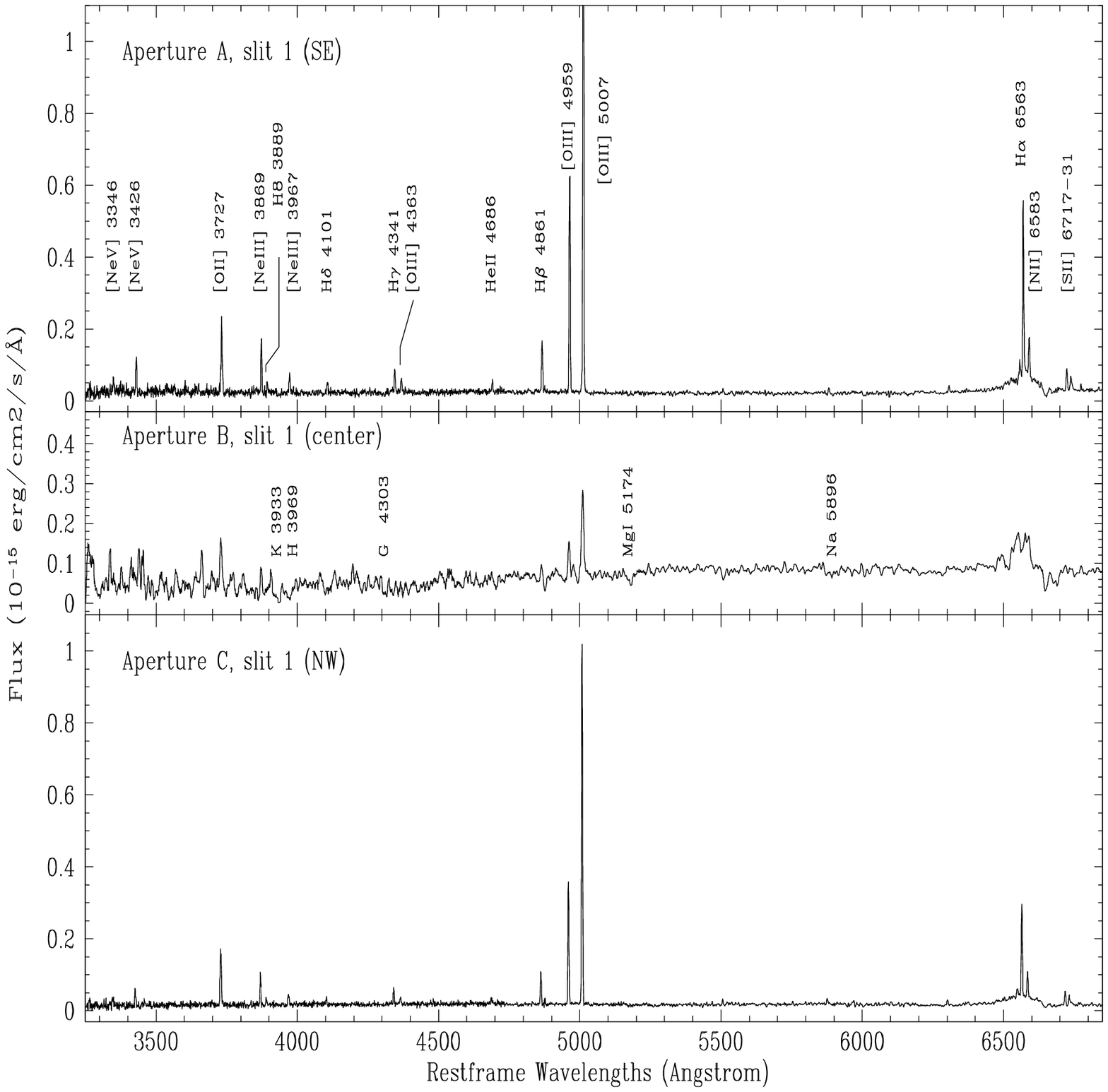}
\caption{Deconvolved spectra of host  galaxy, for the three regions A,
B, C, defined  in Fig.~\ref{slits}.  The size of  the central aperture
(B)  is 2\arcsec,  while apertures  A and  C, that  correspond  to the
external parts of the galaxy, are 4\arcsec\ wide.}
\label{spec1Dzoom}
\end{figure*}

Additional data  were obtained  with FORS2 at  VLT/UT4 YEPUN,  on the
night of  May 13,  2002, with  the slit oriented  to include  both the
quasar  host and  the companion  galaxy G1.   The PA  of this  slit is
$-41^{\circ}$ ({\it  slit~2} in  Fig.~\ref{slits}).  As for  the FORS1
observations,  the MOS configuration  was chosen  in order  to observe
several  PSF stars.   Only one  grism was  used, G600RI,  covering the
spectral range  4560--7310\AA\ in the  rest frame. A single  exposure of
1500 seconds was taken.  The reduction was carried out in the same way
as for the FORS1 observations.  However, in this case, the pixel scale
is  1.5\AA\,  in the  spectral  direction  and  0.252$\arcsec$ in  the
spatial direction.  The resolving power is the same as for FORS1,
$R \sim 700$.

Figure~\ref{extracted_profiles} displays  the intensity profiles  of the
(deconvolved)  host galaxy  of  \obj\  in broad band, along  the  two slit  positions,
showing structures with a typical  size of a few kpc. These structures
are  real and  play in  important role  in the  interpretation  of the
spectra, as they are prominent in the emission lines of the host.

The spectroscopic version of  the MCS deconvolution algorithm was used
to  process  the 2D  spectra,  in the  same  way  as for  HE~1503+0228
(Courbin  et  al.  \cite{Courbin_b}),  leading  to the  decompositions
displayed  in  Figs.~\ref{spec1D}  and \ref{spec1Dzoom}.  While  the
quasar spectra  are very  similar in both  slits, the host  shows much
stronger and  more spatially extended  emission lines in {\it  slit~1}
than in  {\it slit~2}.  

\section{The ISM of the host}
\label{ISM}

The  ISM of the  host of  \obj\ is  particularly highly  ionized.  The
spectra  extracted in  Fig.~\ref{spec1Dzoom}, e.g.,  along {\it
slit~1} not only  show prominent [\ion{O}{iii}] emission lines, but
also  \ion{He}{ii}, [\ion{Ne}{iii}] and  [\ion{Ne}{v}]
emissions.    The   ionization  potentials   of   these  species   are
respectively 54.6,  40.96 and  97.26~eV, i.e.\  1.5 to 3  times higher
than the  already high  ionization potential of [\ion{O}{iii}] (35~eV).  Line ratios  measured at different locations along  the slit can
be used  to understand the physical  processes at work in  the host of
\obj.   We have  measured  them for  all  significant emission  lines,
relative to H${\beta}$, as is summarized in Table~\ref{flux}.

\begin{table}[t]
\centering
\caption[]{Line intensities relative to H${\beta}$, in apertures A, B
and C of {\it slit~1},  and integrated in a 6\arcsec\ aperture in {\it
slit~2}, that avoids the companion galaxy G1.}
\begin{tabular}{lccccc}
\hline\hline  Line  &$\lambda(\AA)$   &  A  &  B  &  C  &   Slit  2  \\  
\hline
\ion{Ne}{v}   &(3346)     &    0.26    &--     &0.15    &--    \\
\ion{Ne}{v} &(3426)  &  0.64  &--  &0.57 &  --\\  
\ion{O}{ii}&(3727)  & 1.46 &3.42  &2.06& --\\  
\ion{Ne}{iii} &(3869)  & 0.94&1.47  & 0.86&  --\\ 
H${\mathrm{8}}$  &(3889) &  0.12 &--  &0.21  & --\\
\ion{Ne}{iii}&(3967) & 0.44  &-- & 0.34& --\\ 
H${\mathrm{\delta}}$&(4101)& 0.27  &--&0.17 &  --\\ 
H${\mathrm{\gamma}}$ &(4341)&  0.46 &--&0.34  &  --\\  
\ion{O}{iii}  &(4363)  &  0.32  &--  &0.29  &  --\\
\ion{He}{ii}&(4686) &  0.21 &-- &0.22  & --\\ 
H${\mathrm{\beta}}$&(4861) &  1.00  &1.00 &1.00  &  1.00\\ 
\ion{O}{iii}&(4959) &  4.08 &2.18&3.43 & 2.44\\ 
\ion{O}{iii}  &(5007) & 12.14&6.72 &10.12& 4.40\\
H${\mathrm{\alpha}}$   &(6563)&    4.10   &8.95   &3.88    &   3.38\\
\ion{N}{ii}    &(6583)   &   1.52    &6.32   &1.53    &   1.29\\
\ion{S}{ii}    &(6717)   &   0.44    &0.53   &0.41    &   0.34\\
\ion{S}{ii} &(6731) & 0.37 &0.34 &0.37 & 0.36\\ \hline
\end{tabular}
\label{flux}
\end{table} 

\begin{table}[t]
\centering
\caption{Intensity ratios in apertures A, B  and C of {\it slit~1} and
for {\it slit~2} (s2).}
\begin{tabular}{lcccc}
\hline\hline    Ratio   &   A   &   B   &   C   &s2   \\   \hline  log
([\ion{O}{iii}]$\lambda 5007$)/(H${\beta}\ \lambda 4861$) &1.1 &
0.8&     1.0&0.4\\     log ([\ion{O}{ii}]$\lambda     3727$)/([\ion{O}{iii}]$\lambda   5007$)   &   -0.9&  -0.3&-0.7&--\\   log
([\ion{N}{ii}]$\lambda   6583)$/(H${\alpha}\ \lambda   6563$)
&-0.4&-0.2 &-0.4&-0.4 \\ \hline
\end{tabular}
\label{logs}
\end{table}

\begin{figure}[t]
\centering \includegraphics[width=9cm]{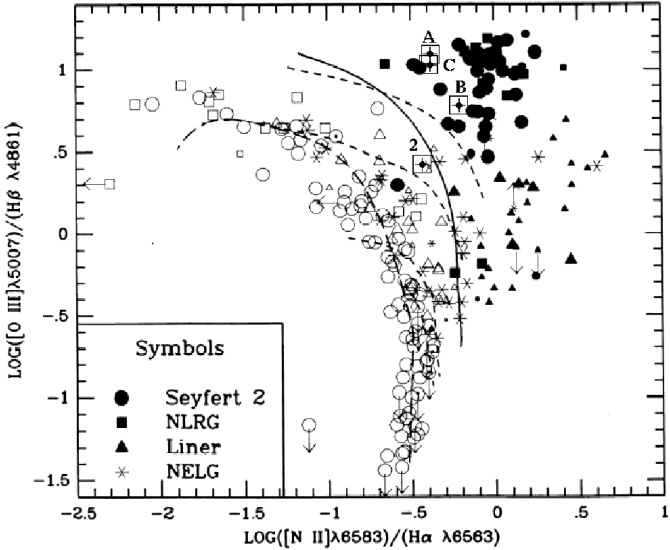}
\caption{Diagnostic    diagram    from    Veilleux   \&    Osterbrock 
\cite{veilleux}, where  the measurements for different  regions of the
host of  \obj\ have been  overplotted. Four points are  available, for
apertures A,  B, C  of {\it  slit~1} and  for {\it  slit~2}.  They are
represented  as  crosses with  associated  1$\sigma$  error box.  Open
symbols,   following   Veilleux   and  Osterbrock   (\cite{veilleux}),
represent typical \ion{H}{ii}  regions.  Black symbols are active
galaxies.   All measurements  of  Veilleux \&  Osterbrock include  the
nucleus and the host of the objects.}
\label{veill}
\end{figure}

Baldwin  et   al.   (\cite{baldwin}),  and   Veilleux  and  Osterbrock
(\cite{veilleux}),  have  introduced  the  so-called  {\it  diagnostic
diagrams} in order to distinguish between several plausible ionization
processes.  Comparing the logarithms of  several line ratios, in  particular those  involving highly  ionized species versus  less  ionized ones,  they have been able to separate  \ion{H}{ii} regions  from  other objects
powered by non-thermal processes.  

The relevant intensity ratios for  the host of \obj\ are summarized in
Table  \ref{logs} and  compared to  the published  sample of  AGNs and
\ion{H}{ii} regions  of Veilleux \&  Osterbrock (\cite{veilleux})
in Fig.~\ref{veill}.   No extinction  correction has been  applied, as
the intensity ratios are for lines at similar wavelengths.  Reasonable
reddening corrections would only lead to negligible changes.

The most striking result is that, while the measurements for {\it slit
2}  show intermediate  ionization  levels, all  three  ratios for  the
extractions along {\it  slit~1} fall well into the  AGN zone, even for
regions  situated as  far  as 6  kpc  from the  central quasar.   This
empirical comparison of the ionization level of the ISM in the host of
\obj\  with the  sample  of Veilleux  \& Osterbrock  (\cite{veilleux})
suggests that at least a significant part of the ionization comes from
non-thermal  processes.   In  addition,  the  significant  differences
between {\it slit~1} and {\it slit~2}, as well as between the external
apertures  (A and  C) and  internal apertures  (B) extractions  of the
spectra, along  {\it slit~1}, show  that the ionization  degree is not
constant throughout  the galaxy.  The  ionization, if produced  by the
central QSO,  might not be isotropic because of shadowing effects on UV radiation from the nucleus. However, it is more likely that density variations in the gas clouds give rise to different ionization degrees, as suggested by Stockton et al. (\cite{stock02}).

The strongest narrow line emissions we detect in the host of \obj\ are
located far away from the nucleus,  at about 5--6~kpc along the direction
of {\it slit~1}.  The spectrum  obtained through {\it slit~2} does not
show such strong emission lines far away from the nucleus.  Making the
analogy  with  Seyfert  galaxies   that  display  radio  jets  may  be
relevant. In some of these galaxies, ionized gas is detected as far as
several   kiloparsecs   away  from   the   center.    Evans  et   al.\
(\cite{evans}),  Wilson  et  al.\  (\cite{wilson}),  Fosbury  et  al.\
(\cite{fosb}), Tadhunter \& Tsvetanov (\cite{tad}) argue that the high
level of ionization  is due both to the UV  radiation from the nucleus
and  to shock  induced ionization  by the  radio jet.   A  more recent
example of shock  induced ionization is presented in  Morganti et al.\
(\cite{morganti03}), who  observed the Seyfert  galaxy IC~5063. Thanks
to the high spatial resolution of their radio observation, the authors
identify a  radio jet, exactly overimposed on  ionization ``hot spot''
seen   in   the   optical   spectrum   of   the   galaxy,   e.g.,   in
[\ion{O}{iii}].

In our  case, the  low resolution radio  map of  Fig.~\ref{nvss} shows
radio jets oriented N-S, i.e.\  nearly exactly in between {\it slit~1}
and {\it slit~2}.   If the emission is induced by a  shock, and if the
shock is along  the direction of the radio jet,  one would thus expect
similar  contributions  in both  slits,  in  sharp  contrast with  the
observations.  VLA or
VLBI observations of \obj\ would help to clarify which influence, if any, the radio jet has on this high ionization.

Extended Emission Line  Regions (EELR) are also found  in RLQ (Boroson
et al.  \cite{boroson85}), that have strong [\ion{O}{iii}](5007\AA) and other highly ionized emissions, no  \ion{Fe}{ii} in their central part (i.e.,
the quasar  spectrum), extended radio lobes, and  steep radio spectra.
\obj\ displays these characteristics,  except for the  last feature
for  which we lack  information.  EELR  are often  spatially unrelated
with  interstellar emissions,  but scales  involved for  the  EELR are
systematically larger (around 70--80~kpc) than ours ( $<$~10~kpc).  In
the studies  of Stockton et al. (\cite{stock02}) and Wilman et al.
({\cite{wilman}), assumption  is made  that direct radiation  from the
central AGN  is the source of  ionization, while the source  of gas is
thought to be recent interaction or merger. This scenario explains why
EELR are not  found in  all RLQ.  The  link between  radio  jets and  EELR
however remains unclear.  EELR where at least two different ionization levels coexist can be explained by density variations (Stockton et al. \cite{stock02}). This seems to be the case for \obj \ .  Whether or not a shock induced ionization exists, direct ionization by the high energy radiation from the AGN,  typical in EELR, even in regions as far as 6 kpc from the nucleus, is sufficient to explain the spectral features of the present object.


\section{Redshift and environment}
\label{neighb}

\subsection{\obj}

Measuring the redshift of  \obj\ on all available  emission lines leads to slightly different estimates
when  using the central  quasar or  its host  galaxy.  We  measure $z=
0.1448 \pm 0.0001$ for the host  only, and $z= 0.1443 \pm 0.0001 $ for
the  quasar only  ($\Delta  v=150~\kms$), or  a  mean redshift  of
$z=0.1445 \pm 0.0003$. The measurement  done on the host's spectrum is
for the center of the galaxy,  hence not affected by rotation, if any.
However, if the velocity field observed for the gas (see next section)
is due to gas inflow/outflow  due to merger activity, this value might
be biased.

\begin{figure}[t!]
\centering \includegraphics[width=9cm,height=5cm]{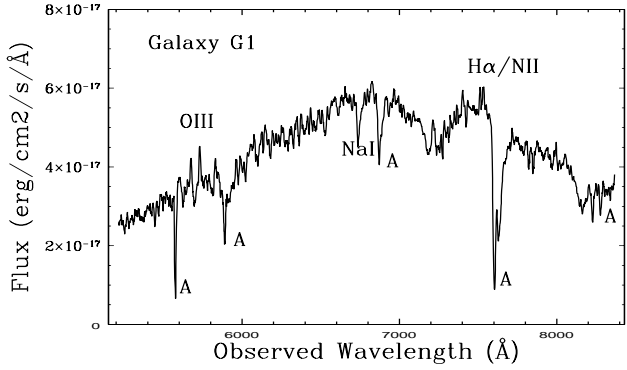}
\caption{Integrated   spectrum    of   companion   galaxy    G1   (see
Fig.~\ref{slits}).   One strong  absorption feature  is  detected: the
sodium  \ion{Na}{i}(5896\AA) line.   All  other absorptions  are
tellurics (labeled  ``A'' on the figure).  The  emission lines visible
here are  extended emissions of the host of  \obj, that are 
contaminating the spectrum of G1.}
\label{compa}
\end{figure}

\begin{figure}[h!]
\centering \includegraphics[width=9cm]{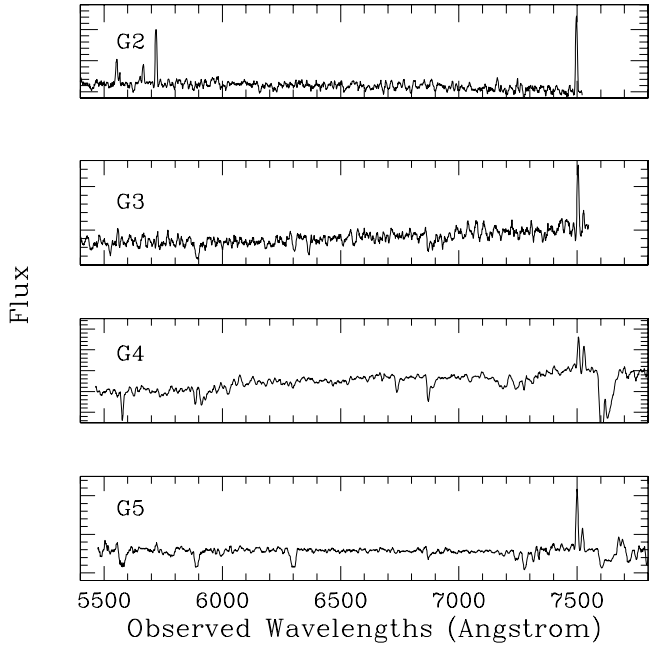}
\caption{Flux calibrated  spectra of 4  neighbours of \obj.   All show
clear H${\alpha}$  and [\ion{N}{ii}]  emissions.  G2 and  G3 show the  [\ion{O}{iii}] and H$\beta$ emission lines as well.}
\label{vois}
\end{figure}

\begin{table}[h!]
\centering
\caption[]{Redshifts of the brightest galaxies in the field of view of
\obj.  Each  galaxy is  indicated in Fig.~\ref{field}. The
coordinates and velocities are relative to the \obj.}
\begin{tabular}{lrrcc}
\hline\hline  Galaxy & x(\arcsec)   & y (\arcsec)    & Redshift & Velocity \\
               &  &  &          & ($\kms$)
\\ \hline
\obj & $+$0.0 & $+$0.0 & 0.1443 & 0 \\ 
G1(companion) & $+$2.6 & $-$3.0 & 0.1428 & $-$450\\ 
G2 & $-$25.0  & $-$38.4   & 0.1426 & $-$526\\ 
G3 & $-$54.0  & $-$63.4   & 0.1436 & $-$210\\ 
G4 & $-$9.2   & $-$63.0   & 0.1437 & $-$180\\ 
G5 & $-$110.0 & $+$52.0   & 0.1429 & $-$420\\

\hline
\end{tabular}
\label{red}
\end{table}

\subsection{Environment}

Redshifts were also determined for  5 other objects in the vicinity of
\obj, indicated in Fig.~\ref{field}.  One  of them is galaxy  G1, located 4\arcsec\,  away from \obj\
and present only  in {\it slit~2}.  The  redshift measurement is based
on    the    identification   of    the    sodium   absorption    line
\ion{Na}{i}(5896\AA).   The apparent  emission lines  visible on
this  spectrum are  only due  to  contamination by  the extended  host
galaxy of \obj. The spectrum  of G1 is presented in Fig.~\ref{compa}.
 Measuring  the  redshift  directly  from \ion{Na}{i}  and  other
weaker Fe  lines or by cross-correlating  the spectrum with  that of a
template  spectrum  of  an  elliptical  galaxy  yields  $z=0.1428  \pm
$0.0003.

The redshifts of four  other galaxies, picked-up randomly because they
were in the  unvignetted area of the CCD  chip (Fig.~\ref{vois}), were
also measured,  mainly based on the H${\alpha}$  emission line (Table~\ref{red}).  Galaxy G4 even shows rotation, with a prospected rotation
velocity  of v$_{\mathrm{rot}}\sim80\ \kms$,  measured  1\farcs4 \,
from the center of the galaxy, in the flat part of the rotation curve.
G1, the  closest companion to  \obj, is the  only galaxy out of  the 5
known companions that does not show emission lines.

With at least 6 objects at  almost the same redshift, \obj\ is part of
a small group of  galaxies whose velocity dispersion is $\sigma_v=197~\kms$, with a 95 percent confidence interval of 123~$\kms  \le
\sigma_v \le  480~\kms$.  Considering the group as an isothermal
sphere  with a  mean  radius of  $r=100\arcsec $  (Binney and  Tremaine 
\cite{binney}), its  mass is, $M_{\mathrm{group}}\sim  4\cdot 10^{12}$~M$_{\odot}$.  This places  it among  the loosest  groups  of galaxies,
e.g., when compared with the groups found in the Las Campanas Redshift
Survey (Hein\"am\"aki et al. \cite{heina03}), who find group masses in
the range $M=10^{12}$--$10^{15}$~M$_{\odot}$.


\section{Dynamics of the host galaxy}

Because our observational strategy  is to take the spectra ``on-axis''
rather than a few arcseconds away from the AGN center, the dynamical
information about  the host galaxy  is not lost.   The signal-to-noise
ratio of the data is sufficient to perform quantitative measurement of
the stellar and gas velocity fields.

\subsection{Stellar dynamics}

The  Calcium  \ion{Ca}{ii} absorption  doublet  is  detected  in the  spectrum
obtained in April 2000. It  is outside the accessible wavelength range
in  the  observations   taken  in  May  2002.   Because   of  the  low
signal-to-noise in these  lines, we can not perform  a full extraction
of the stellar  velocity field.  We can however  get significant hints
about  the stellar rotation/motion,  by dividing  the galaxy  into two
parts along the spatial direction, and by cross-correlating them.  Two
spatial extractions  were done along {\it  slit~1}.  The  light of the
quasar host was integrated in  two ``boxes'' lying between 0\farcs8 and 2\farcs8 away  from the center of the  galaxy.  Their cross-correlation,
displayed in  Fig.~\ref{correl} shows  a sharp peak centered  at $v=2
\pm 10~\kms$.  Similar results are found on the {\it slit~2} spectrum.
We can therefore conclude that  stars do not display a global rotation
in the host of \obj.

\begin{figure}[t]
\centering 
\includegraphics[width=8cm,height=5cm]{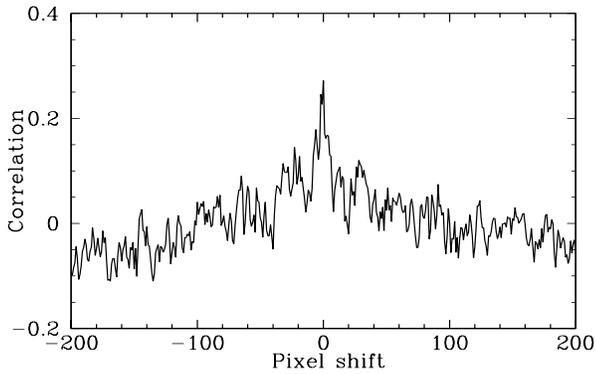}
\caption{Correlation  function of  the  two extracted  regions of  the
host's  spectrum (see  text) along  {\it slit  1}.  Only  the spectral
regions with no emission lines are considered, i.e., from 5246~\AA\ to
5706~\AA, in  the rest frame.  The centroid of  the correlation peak is
located at  0.03 $\pm$ 0.1 pixels,  which translates into  $v=2 \pm 10~\kms$, hence confirming that the stars in the host are not affected by
any circular motion.}
\label{correl}
\end{figure}

\subsection{Gas dynamics}

Contrary to the absorption lines, an obvious velocity field is seen in
the emission lines of the host  galaxy. An example of emission lines is
shown in Fig.~\ref{extr}, where the continuum of  the host galaxy has
been removed.  The gas seen  through the two slit orientations behaves
differently. While {\it  slit~1} reveals the gas  motion, the emission
lines seen along {\it slit~2} remain perfectly straight.

The spatial  profiles in the lines  in Fig.~\ref{extr}  show, as said
previously, that the  gas is not distributed uniformly  in the galaxy.
In  particular, the  [\ion{O}{iii}] emission  line is  composed of  two bright
spots located  2\farcs2\ and 2\farcs4\ away from  the nucleus, and
that correspond perfectly to the  two bright spots already seen in the
spatial  profiles  of  Fig.~\ref{extracted_profiles}.   They  are  not
deconvolution artifacts.  In fact, they are detectable in the original
data without  subtracting the quasar  (Fig.~\ref{profile}). 

The velocity curves have  been determined from several emission lines,  extracted from reduced spectra to avoid artifacts deconvolution could bring, specially near the nucleus. The extraction method is described in Courbin et al. (\cite{Courbin_b}). 
These curves are displayed in  Fig.~\ref{rotcurve}. While there is no velocity
field along {\it slit~2}, the curve  for {\it slit~1} is very sharp in
the  central kiloparsec, and  almost flat  in the  outer parts  of the
galaxy. This is hard to reconcile with models involving pure rotation,
but  we nevertheless  attempted  to  follow the  same  approach as  in
Courbin   et  al.   (\cite{Courbin_b}),   for  the   spiral  host   of
HE~1503+0228.

A mass model  is assumed for the galaxy, including  a rotating disk, a
central point  mass and a dark  matter halo.  This mass  model is then
used to  predict the velocity  of the galaxy  at a given point  of the
slit, taking into account the  inclination of the disk and convolution
by  the seeing profile,  using the  known spectrum  of the  PSF.  Many
different initial conditions were used when fitting the model, with no
successful solution.  In fact, we found no way to model simultaneously
the sharp central velocity trough,  that requires a very large central
mass, and the flat external  parts, that require large amounts of dark
matter,  while keeping the various parameters at reasonable values.

\begin{figure}[t!]
\centering                           
\includegraphics[width=4.3cm]{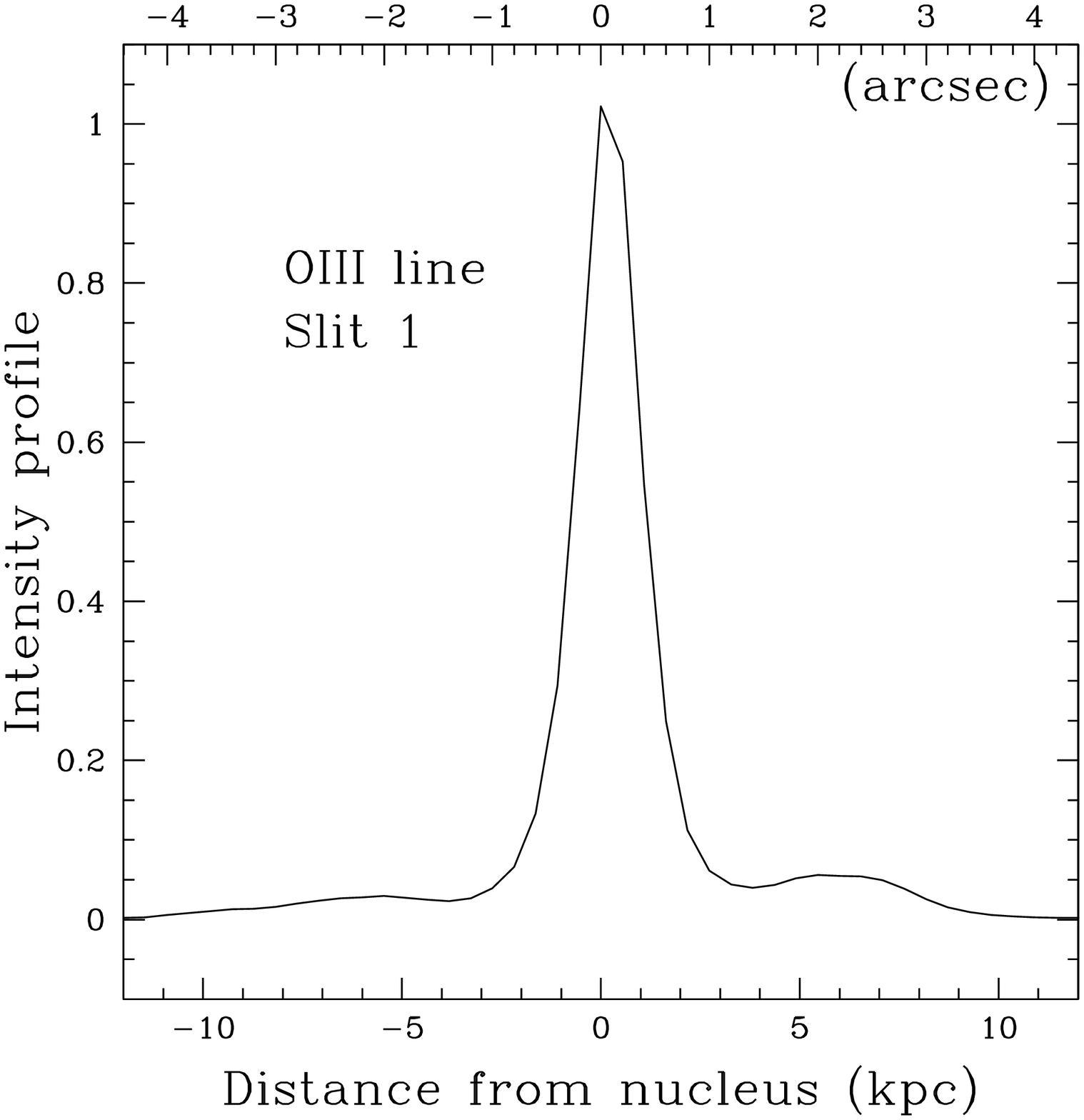}
\includegraphics[width=4.3cm]{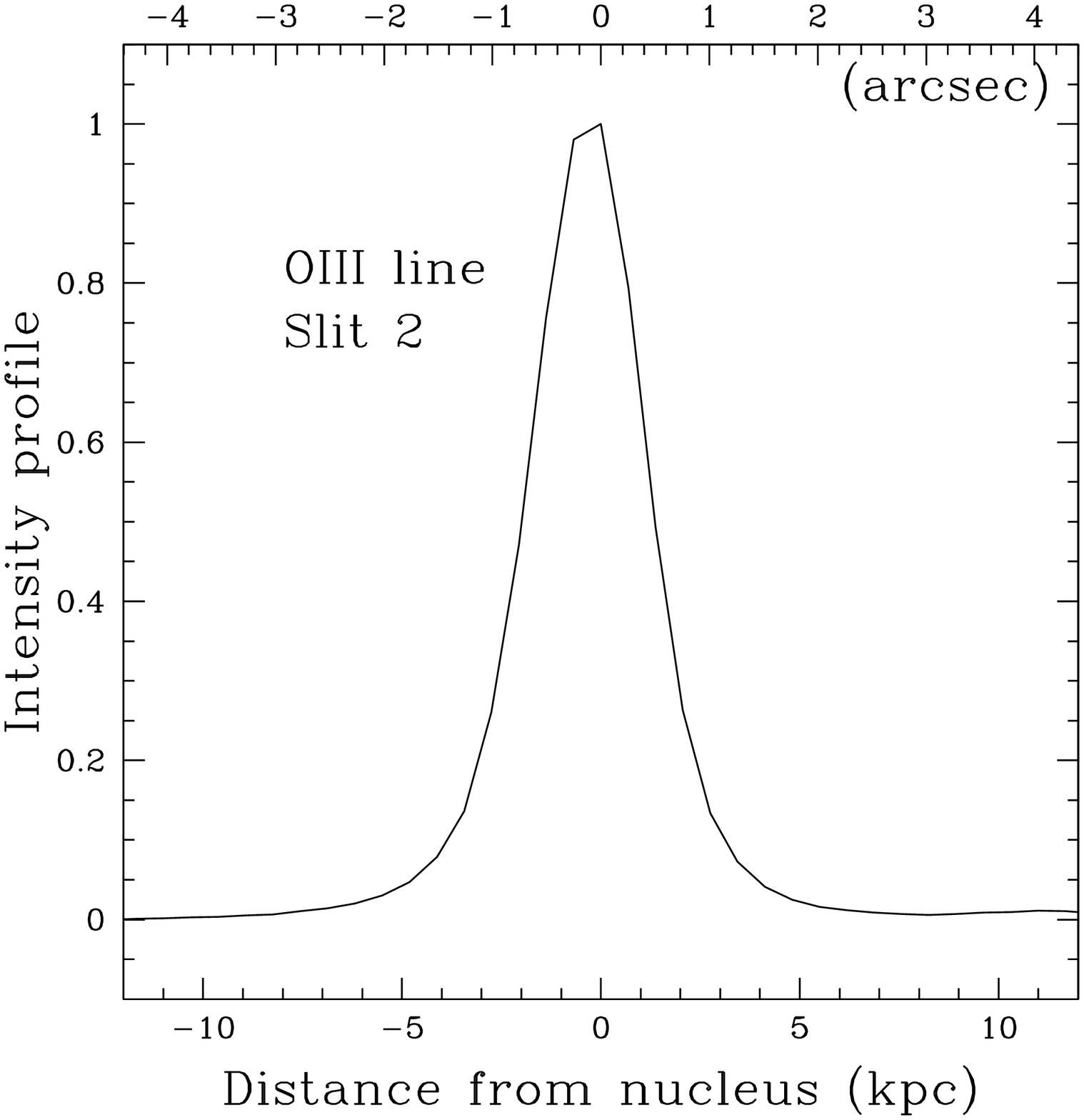}
\caption{Spatial profile of the [\ion{O}{iii}]  emission line for {\it slit~1}
and {\it slit  2} prior to any host/nucleus  decomposition. The dip in
intensity  profile  through {\it  slit  1},  3\arcsec\  away from  the
quasar,  and the  two  ``hot spots''  of  Fig.~\ref{extr} are  already
visible in this raw undeconvolved intensity profile.}
\label{profile}
\end{figure}

We have thus  considered other mass distributions, such  as a modified
Hubble  profile,  power-law  density  profile or  isochrone  potential
(Binney  \&   Tremaine  \cite{binney}),  all   encountering  the  same
difficulty. In the best fits, central   masses  of  the  order   of  $M\sim  10^{10}$~M$_{\odot}$ are always found.   For comparison, the central mass found
for HE~1503+0228 was $M\sim  5\cdot 10^{7}$~M$_{\odot}$, three orders
of magnitude smaller.  Other  groups have estimated the central masses
in galaxies,  most of  the time for  isolated ellipticals like  in the
SAURON  survey (e.g., NGC  3377, Copin  et al.   \cite{copin03}), more
massive ellipticals (M87, Macchetto et al.  \cite{macchetto97}, Cen A,
Marconi  et  al.   \cite{marco01}),  or  Seyfert  galaxies  (NGC~4041,
Marconi  et al.  \cite{marco03};  Mrk 110,  Kollatschny \cite{kola03}).
The masses  they find,  either using the  black hole  central velocity
dispersion relation,  or direct dynamical  modeling, is always  in the
range  $M= 10^{7}$--$10^{8}$~M$_{\odot}$,  with some  supermassive black
holes  reaching up  to  10$^{9}$~M$_{\odot}$,  e.g.,  in the  Seyfert
NGC~5252  (Macchetto \cite{macchetto03}).  The central  mass  in \obj,
assuming gas rotation,  is still an order of  magnitude above the most
massive black holes found in other galaxies.

\begin{figure*}[p!]
\centering           
\includegraphics[width=5.5cm]{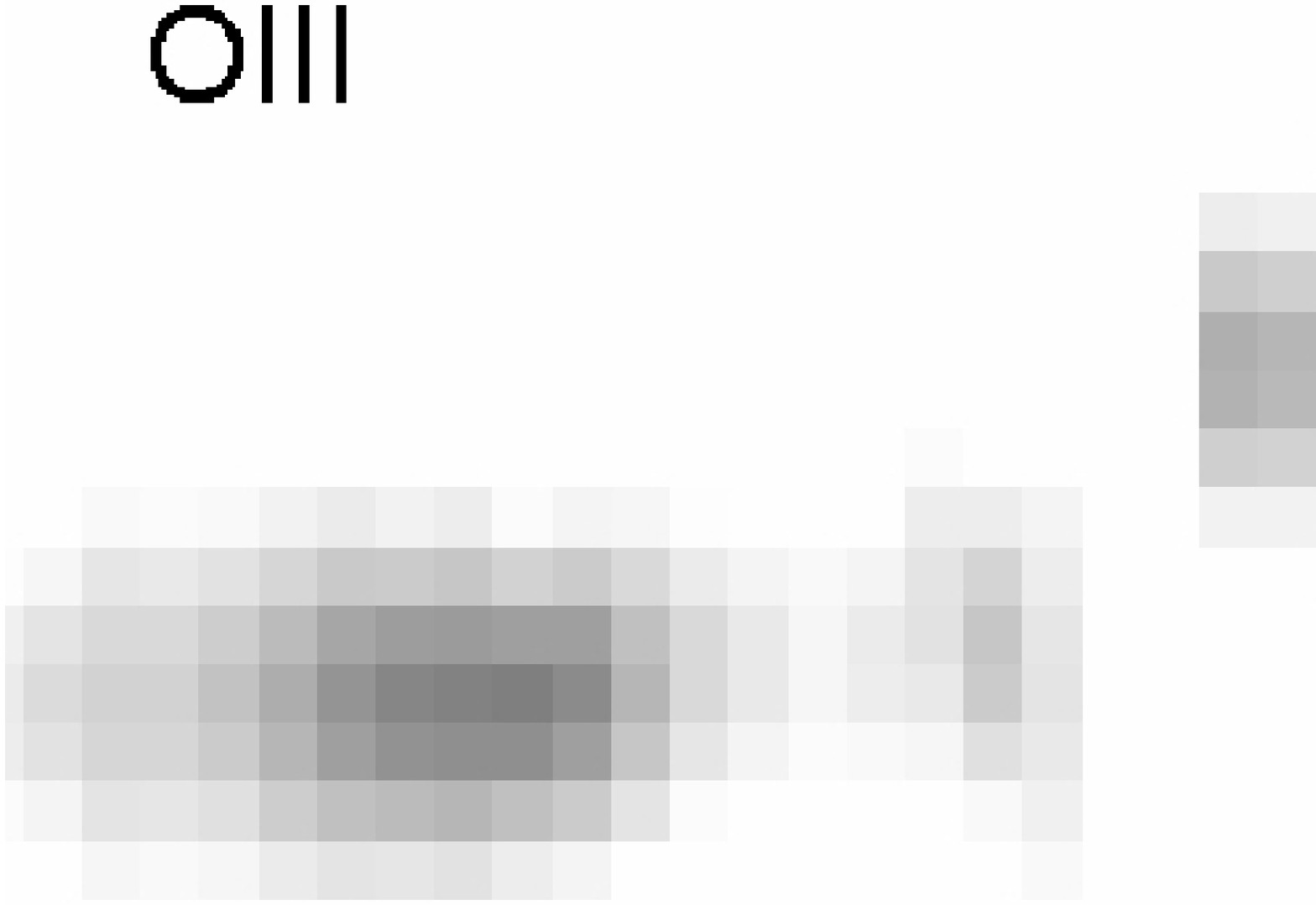}
\includegraphics[width=5.5cm]{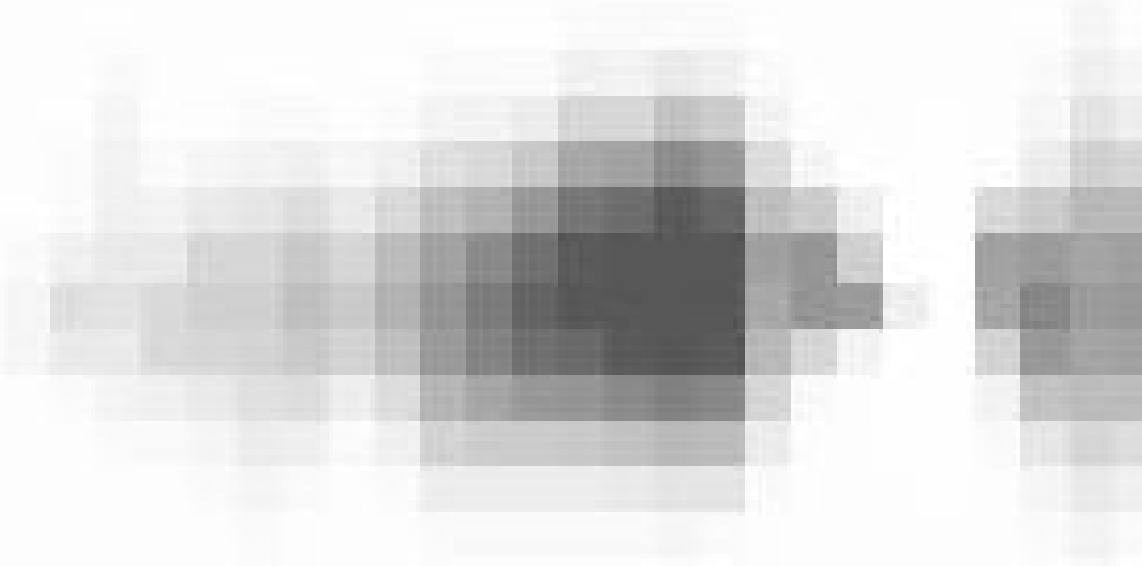}
\includegraphics[width=6.cm]{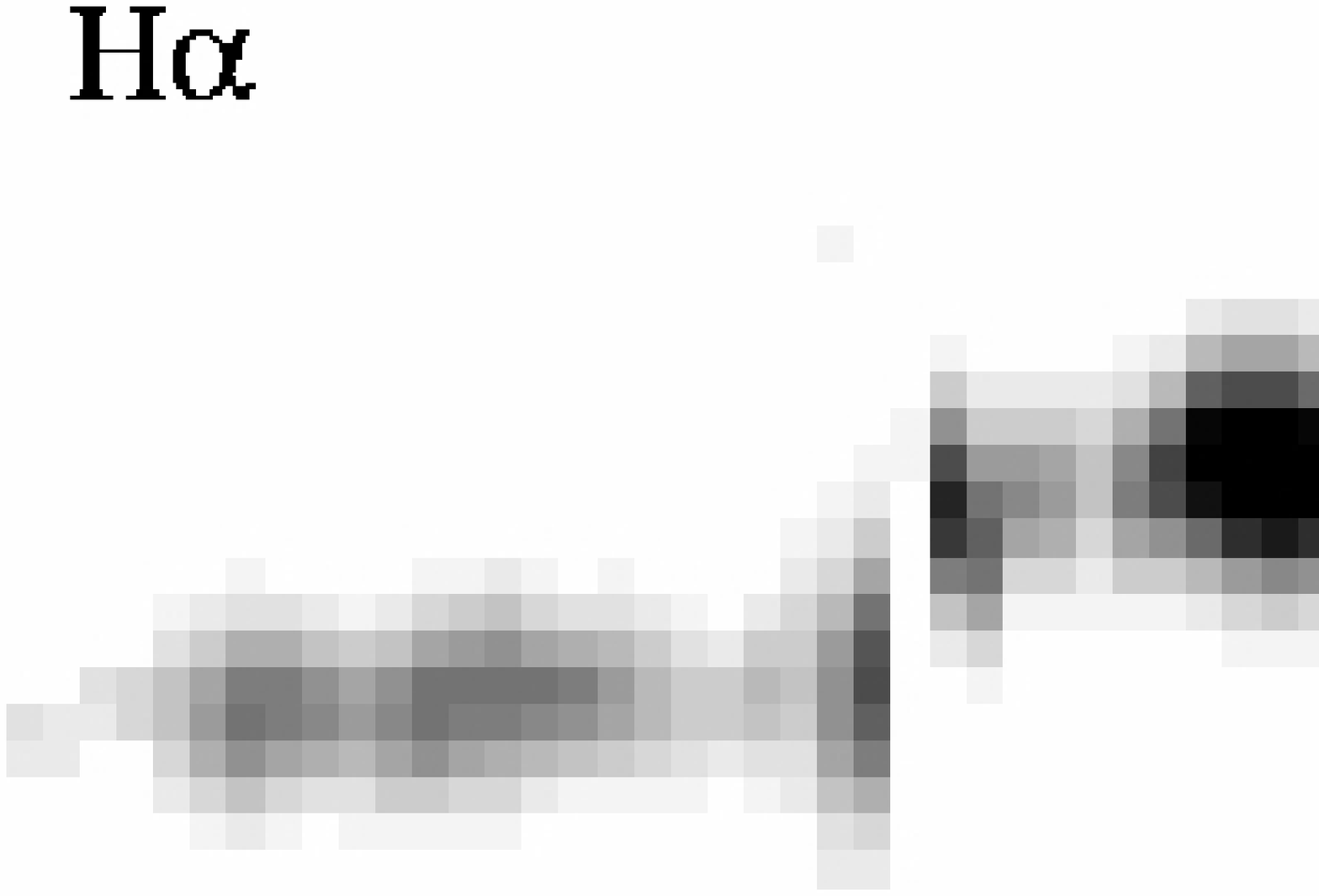}
\includegraphics[width=5.8cm]{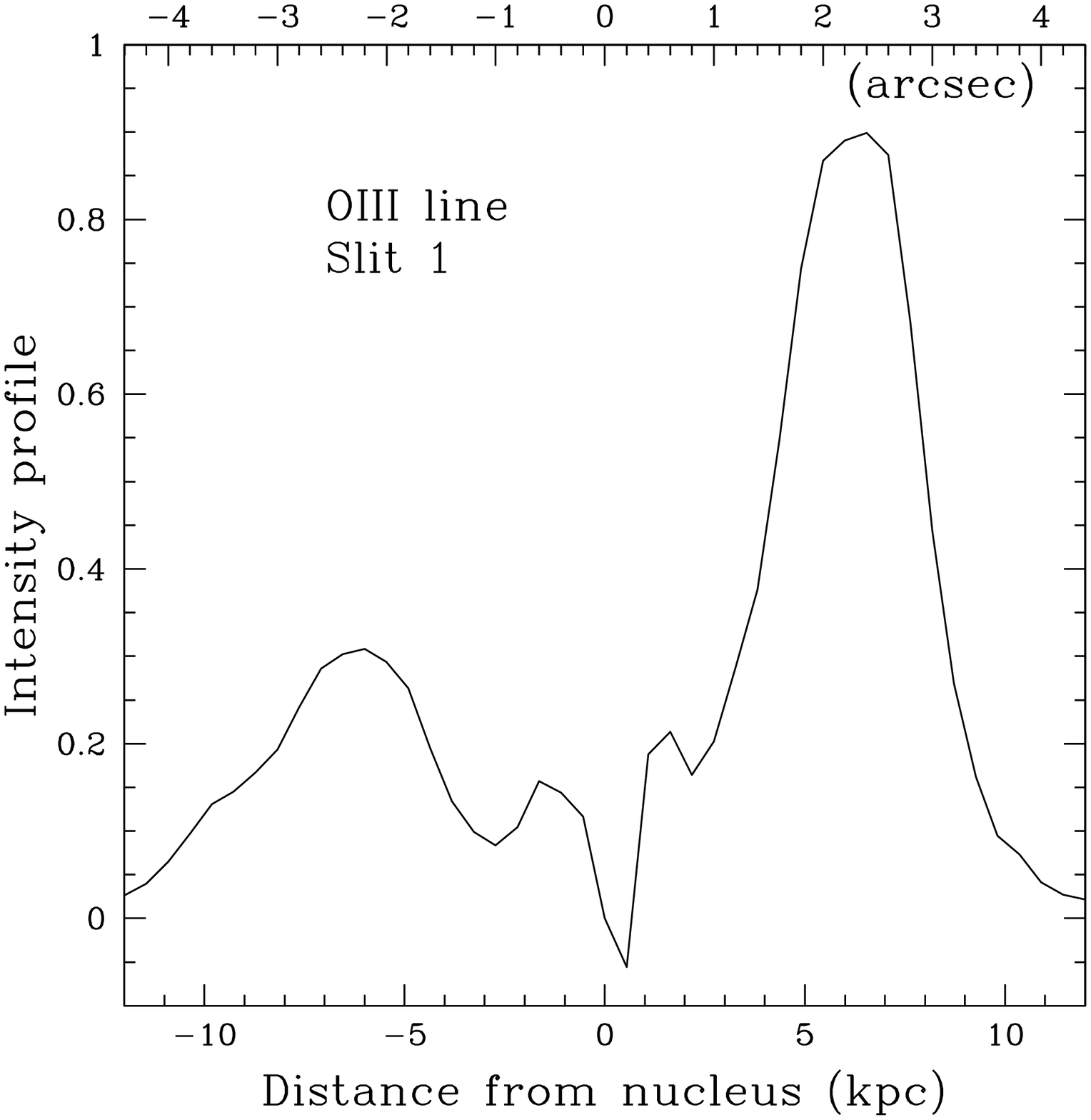}
\includegraphics[width=5.8cm]{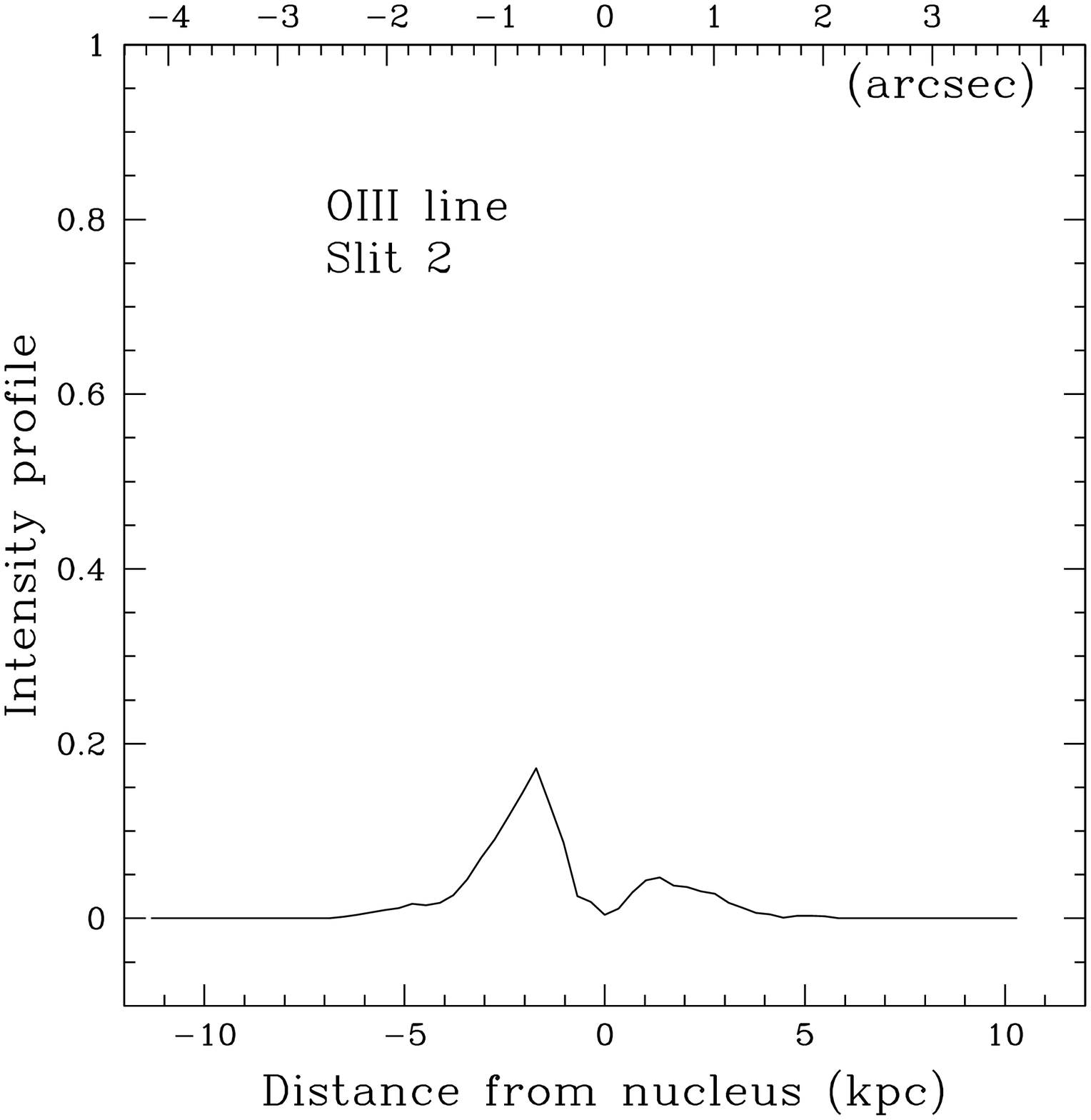}
\includegraphics[width=5.8cm]{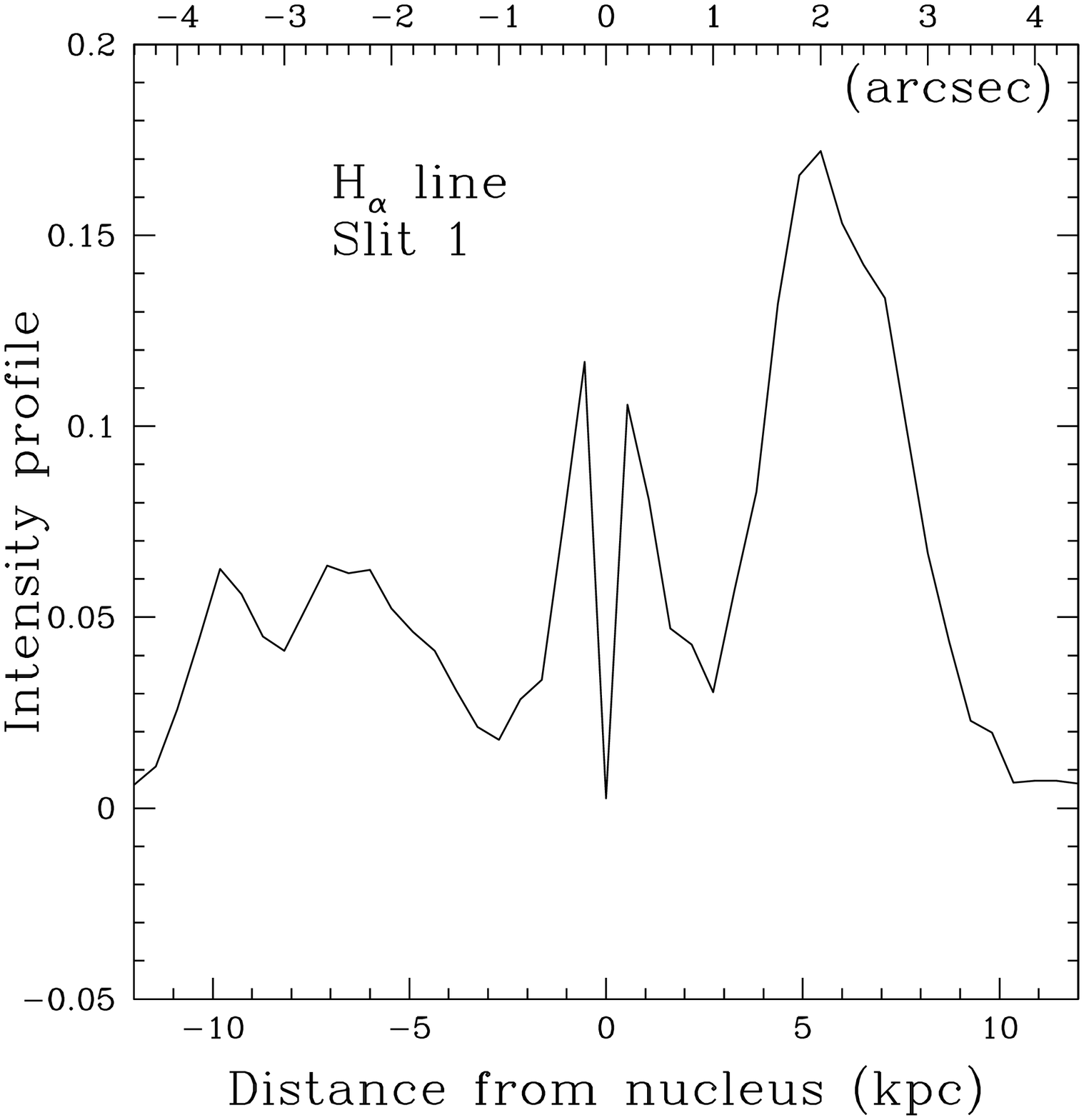}
\caption{{\it Top:} zoom on the [\ion{O}{iii}] (5007\AA) emission
line in  {\it slit~1} (left),  in {\it slit~2}  (middle). The H$\alpha$
emission lines is displayed on  the right panel. The spatial direction
is horizontal and the  wavelength direction vertical.  Both the quasar
spectrum   and  the   continuum   of  the   host   galaxy  have   been
subtracted. The sharp  structures in the very center  (central kpc) of
the galaxy  corresponds to  oversubtraction of  the quasar.   The gas
motion is  striking in {\it slit  1}, while the stars  are not showing
any  motion in  this same  slit position.   No gas  velocity  field is
detected  for {\it slit  2}.  {\it  Bottom: }  Corresponding intensity
profiles.}
\label{extr}
\vspace*{2mm}
\includegraphics[width=8.5cm]{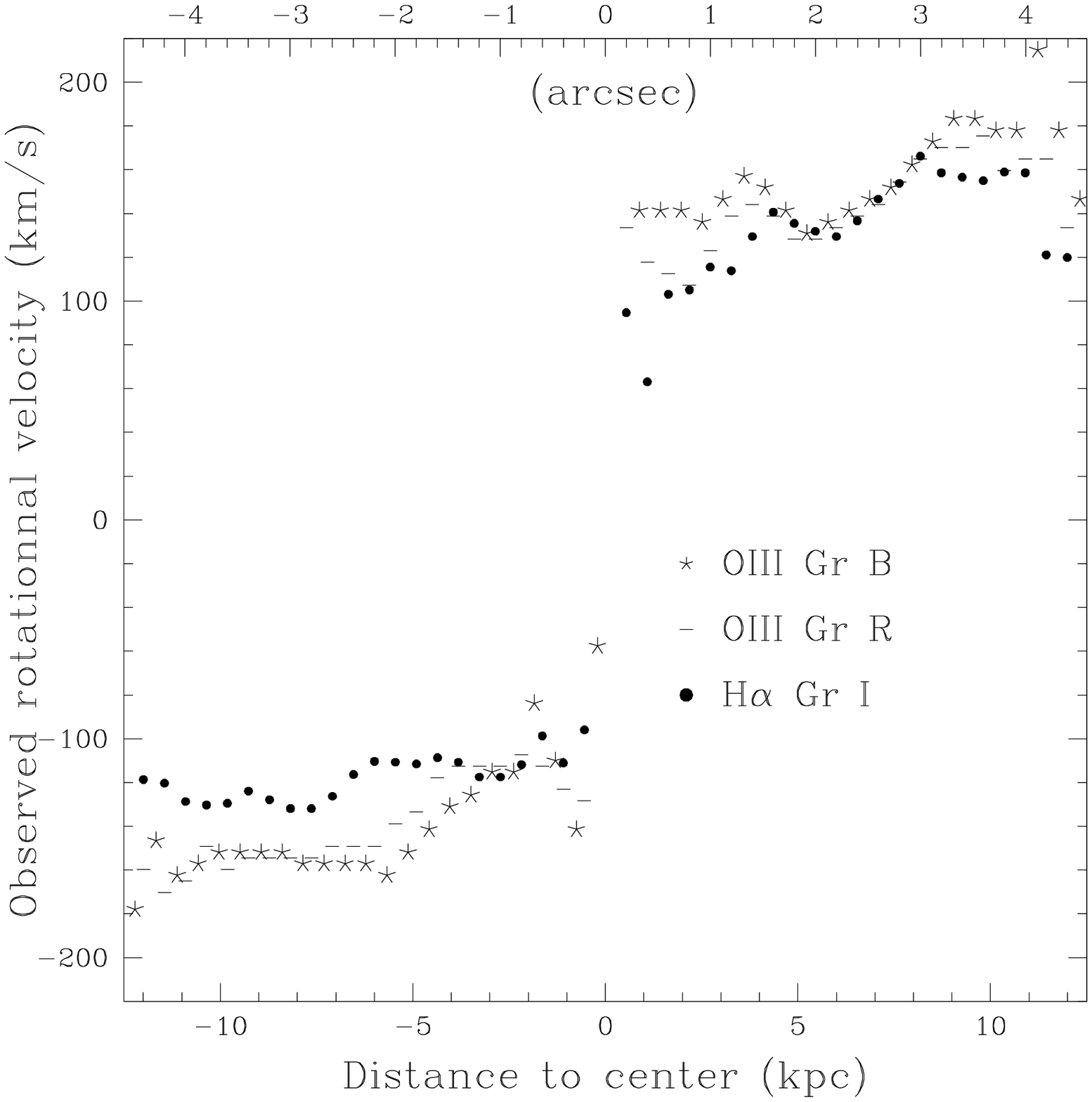}
\includegraphics[width=8.5cm]{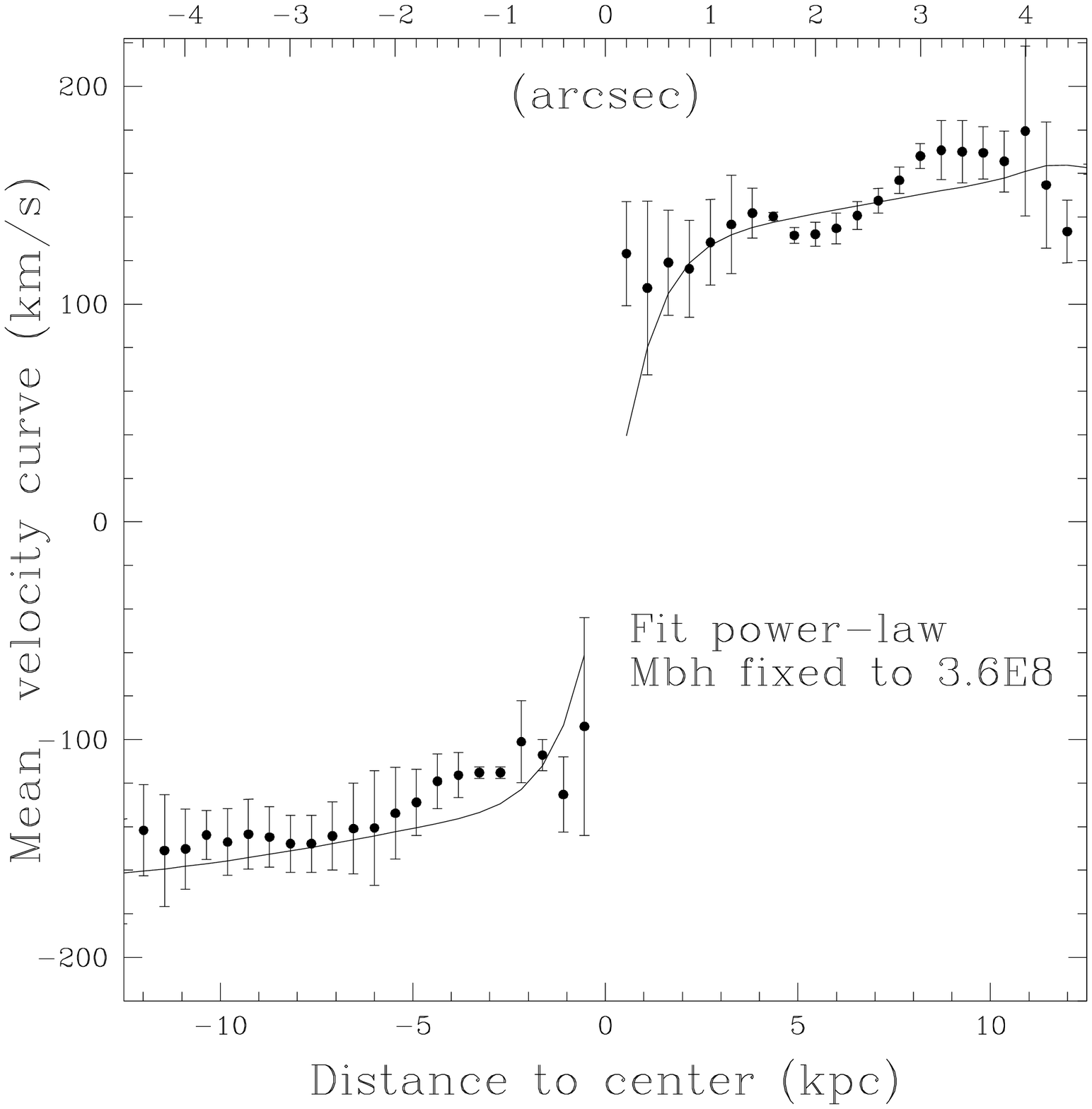}
\caption{{\it Left:} velocity curves obtained for the [\ion{O}{iii}] doublet and
for  the   H$\alpha$  emission  line.   {\it   Right:}  mean  velocity
curve. The data points are the mean of the two [\ion{O}{iii}] measurement and of
the H$\alpha$ measurement. The error bars correspond to the dispersion
between the  three curves of  the left panel.   Our best fit  model is
overplotted.  With a central mass fixed to $3.6~10^8~$M$_{\odot}$, the $\chi^2$  of  the fit   is  acceptable only for model with mass components unrealistically concentrated near the nucleus (see text).}
\label{rotcurve}
\end{figure*}

The  central  mass can  be  estimated  independently  using the  known
correlation between  R$_{\mathrm{BLR}}$, the radius of  the Broad Line
Region (BLR),  and the monochromatic AGN continuum  luminosity at 5100
\AA   \,  (Kaspi   et  al.    \cite{kaspi}).    This  luminosity-based
R$_{\mathrm{BLR}}$ is combined with  a measurement of the BLR velocity
extracted  from the  Full Width  at  Half Maximum  of the  H${\beta}$
emission  line.  Assuming  that  the  motion  in the  BLR  in  AGN  is
virialized, one can use the simple Keplerian relation
\begin{equation}
M_{\mathrm{BH}}=R_{\mathrm{BLR}}\, v_{\mathrm{BLR}}^{2}\, G^{-1},
\end{equation}
to estimate the mass of the  central black hole. In the case of \obj\,
we find $M_{\mathrm{BH}}\sim $3.6 10$^{8}$~M$_{\odot}$.   Again, this is
two  orders of  magnitude lower  than  obtained with  our galaxy  mass
models. 

We have tried to fit the velocity curves with models including a central mass fixed at the above value, i.e. $M_{\mathrm{BH}}=3.6\,10^8$~M$_{\odot}$. We can obtain rather reasonable fits (see Fig. \ref{rotcurve}), but at the expense of completely unrealistic parameters (e.g., a disk with a huge mass density confined to a few parsecs). Indeed, the lower $M_{\mathrm{BH}}$ is always compensated by other mass components which become concentrated near the nucleus.
Nevertheless, all models  lead to a total mass  of $\sim 9.2\,10^{10}$~M$_{\odot}$, within a radius of 10~kpc.

\section{Discussion -- Conclusions}
\label{discu}

In the framework of a comprehensive spectroscopic study of quasar host
galaxies,  we have obtained  VLT optical  observations of  the $z=0.144$
quasar  \obj,  that retained  our  attention  and triggered  follow-up
observations.  Two slit orientations were used (see Fig.~\ref{slits})
and the data  were spatially deconvolved, resulting in  spectra of the
host galaxy decontaminated from the light of the central quasar.

We found the host galaxy of \obj\ is a rather peculiar object.  Simple
observational facts are:

\begin{enumerate}

\item Optical  and near-IR  imaging of \obj\  reveals  an elliptical
morphology,  with  colors bluer than inactive elliptical  galaxies,
indicating a contribution of a young stellar population.

\item The galaxy harbours a  bright quasar, whose radio loudness index
places it  at the limit between  radio quiet and  radio loud. Extended
lobes are seen in the radio up to 5\arcmin\, away from the quasar.

\item The ISM of the  host is heterogeneous and highly ionized  by the central AGN. Ionization triggered by the radio jet can not be excluded.

\item While the  gas emission lines observed through  {\it slit~1} all
show a clear velocity field,  the stellar absorption lines do not display any sign of rotation, in accordance with the observed elliptical morphology.

\item Neither the  stars nor the gas are affected  by a velocity field
in the  {\it slit~2}  observations, almost perpendicular to  {\it slit
1}.

\item No mass model involving rotation  of a gas disk fit properly the
gas velocity  of {\it slit  1} unless invoking extremely  high central
masses for  the galaxy an  order of magnitude  above that of  the most
massive black holes known ($> 10^9$~M$_{\odot}$).

\item \obj\ is part of a small group with $\sigma_v\sim 197~\kms$ or
M$_{\mathrm{group}} \sim 4\cdot 10^{12}$~M$_{\odot}$.

\item The  closest companion  G1, of \obj\  is located  4\arcsec\ away
from it, with a velocity difference of 450~$\kms$. No trace of gas emission is detected.

\end{enumerate}

In order  to explain  the main characteristics  of the host  galaxy of
\obj\ and environment, we propose the following interpretation.

The motion of  the gas, as deduced from the  emission lines, cannot be
interpreted as Keplerian motion in any reasonable mass model.  It must
be affected by  some interaction with an intervening  object. At least
two possibilities  arise: (1) the dynamics  of the gas  is modified by
interaction with  matter and/or radiation from  the AGN or  (2) it has
been affected  by a recent  collision with the neighbouring  galaxy G1.
As the central AGN of \obj\  does not appear very different from other
quasars in the sample,  which display relatively mundane host galaxies
in which  the ISM does  not show obvious  signs of disturbance  by the
central AGN (e.g., HE\,1503+0228, Courbin et al \cite{Courbin_b}), we consider
the second explanation as most likely.

We propose that the ISM of  both galaxies (host and G1) has been swept
out during a recent close interaction and has had insufficient time to
relax.  Using the  velocity dispersion in the group  as an estimate of
the  relative velocity of  the two  interacting galaxies,  we estimate
that  it would  have taken  $\sim  3 \times  10^7$~years  for the  two
galaxies to  reach their present  projected separation.  This  is much
smaller  than the  typical relaxation  time of  a merger  ($\sim 10^9$~years, Wright 1990, Barnes 1989).

Now, we need to explain  why, although quasars of similar luminosities
(e.g., HE\,1503+0228) seem unable to  ionize the interstellar gas outside the
central AGN  region, the central engine  of \obj\ is  able to strongly
excite  and ionize  the  ISM  as far  as  5 to  6  kiloparsec. If  the
collision with galaxy G1 has been  strong enough to remove the gas out
of the host galaxy, even in  the most central regions, nothing is left
to absorb the  ionizing radiation of the quasar which  is then able to
propagate  freely to  large distances,  until it  hits the  gas clouds
moving far away from the center and strongly ionizes them.

In our  opinion, a  recent collision  with galaxy G1  is thus  the key
factor allowing to explain both the peculiar motion of the gas and its
high degree  of ionization by the  central quasar, even  very far away
from the source of radiation.

\begin{acknowledgements}
The  authors would  like to  thank Fran\c{c}oise  Combes  and Raffaela
Morganti   for   useful  discussions   and   suggestions.   The   NASA
Extragalactic    Database    (NED)   has    been    used   for    this
research. G\'eraldine Letawe is  a teaching assistant supported by the
University of  Li\`ege, (Belgium).  Fr\'ed\'eric  Courbin acknowledges
financial  support from  the European  Commission through  Marie Curie
grant  MCFI-2001-0242.   The  P\^ole d'Attraction  Interuniversitaire,
P5/36 (PPS Science Policy, Belgium) is also thanked.
\end{acknowledgements}

\end{document}